\documentclass[aps,10pt,notitlepage,twocolumn,nofootinbib,superscriptaddress]{revtex4-2}

\usepackage{graphicx,color}
\usepackage{amsmath}
\usepackage{amssymb}
\usepackage{hyperref}
\usepackage[utf8]{inputenc}
\usepackage[english]{babel}
\usepackage{epsfig}
\usepackage{subfigure}
\usepackage{wasysym}
\usepackage{color,xcolor}
\usepackage{amsmath}
\usepackage{bm}
\usepackage{epsfig}
\usepackage{amsfonts}
\usepackage{dcolumn}
\usepackage{float}

\usepackage{comment}

\hypersetup{colorlinks=true, linkcolor=blue, citecolor=green}

\usepackage{dcolumn}
\usepackage{bm}
\usepackage{ifpdf}
\usepackage{hyperref}
\usepackage{float}
\usepackage{bm}
\usepackage{xcolor,color,graphicx,graphics}
\usepackage[OT1]{fontenc}
\usepackage{latexsym,amssymb,amsmath,amsfonts}
\usepackage{makeidx}
\usepackage{epsfig}
\usepackage{epstopdf}
\usepackage{mathrsfs}
\hypersetup{colorlinks=true, linkcolor=blue, citecolor=green}
\usepackage{enumerate}
\usepackage{xcolor}
 \usepackage{multirow}

\begin{document}

\title{Dyonic regular black bounce solutions in General Relativity}

	\author{Ednaldo L. B. Junior} \email{ednaldobarrosjr@gmail.com}
    \affiliation{Faculdade de F\'{i}sica, Universidade Federal do Pará, Campus Universitário de Tucuruí, CEP: 68464-000, Tucuruí, Pará, Brazil}
\affiliation{Programa de P\'{o}s-Gradua\c{c}\~{a}o em F\'{i}sica, Universidade Federal do Sul e Sudeste do Par\'{a}, 68500-000, Marab\'{a}, Par\'{a}, Brazill}

     \author{José Tarciso S. S. Junior}
    \email{tarcisojunior17@gmail.com}
\affiliation{Faculdade de F\'{i}sica, Programa de P\'{o}s-Gradua\c{c}\~{a}o em F\'{i}sica, Universidade Federal do Par\'{a}, 66075-110, Bel\'{e}m, Par\'{a}, Brazill}

	\author{Francisco S. N. Lobo} \email{fslobo@ciencias.ulisboa.pt}
\affiliation{Instituto de Astrof\'{i}sica e Ci\^{e}ncias do Espa\c{c}o, Faculdade de Ci\^{e}ncias da Universidade de Lisboa, Edifício C8, Campo Grande, P-1749-016 Lisbon, Portugal}
\affiliation{Departamento de F\'{i}sica, Faculdade de Ci\^{e}ncias da Universidade de Lisboa, Edif\'{i}cio C8, Campo Grande, P-1749-016 Lisbon, Portugal}

    \author{\\Manuel E. Rodrigues} \email{esialg@gmail.com}
\affiliation{Faculdade de F\'{i}sica, Programa de P\'{o}s-Gradua\c{c}\~{a}o em F\'{i}sica, Universidade Federal do Par\'{a}, 66075-110, Bel\'{e}m, Par\'{a}, Brazill}
\affiliation{Faculdade de Ci\^{e}ncias Exatas e Tecnologia, Universidade Federal do Par\'{a}, Campus Universit\'{a}rio de Abaetetuba, 68440-000, Abaetetuba, Par\'{a}, Brazil}

     \author{Luís F. Dias da Silva} 
        \email{fc53497@alunos.fc.ul.pt}
\affiliation{Instituto de Astrof\'{i}sica e Ci\^{e}ncias do Espa\c{c}o, Faculdade de Ci\^{e}ncias da Universidade de Lisboa, Edifício C8, Campo Grande, P-1749-016 Lisbon, Portugal}

    \author{Henrique A. Vieira} \email{henriquefisica2017@gmail.com}
\affiliation{Faculdade de F\'{i}sica, Programa de P\'{o}s-Gradua\c{c}\~{a}o em F\'{i}sica, Universidade Federal do Par\'{a}, 66075-110, Bel\'{e}m, Par\'{a}, Brazill}

\begin{abstract}

This work explores dyonic black bounce (BB) solutions within the framework of General Relativity (GR), coupled with nonlinear electrodynamics (NLED) and scalar fields (SFs). Previous research has employed NLED and SFs to obtain BB solutions in GR; however, these solutions typically assume the presence of either magnetic monopoles or electric charges exclusively as components of the Maxwell-Faraday tensor.
In this study, we examine static and spherically symmetric BB solutions that incorporate both magnetic and electric components, forming what are known as dyon solutions. A dyon is a particle characterized by the coexistence of both magnetic and electric charges. We determine the NLED Lagrangian density and the scalar field potential that produce these solutions and analyze the associated gravitational configurations, focusing on horizons, the behavior of the metric function, and spacetime regularity as described by the Kretschmann scalar.
Notably, we present the first BB solution derived from the coupling of a linear electromagnetic Lagrangian and a scalar field with an associated potential as the matter source. This work broadens the class of non-singular geometries in the literature and opens new avenues for investigating dyonic BB solutions within the context of other modified gravity theories.

\end{abstract}
\date{\today}
\maketitle
\def\HMS{{\scriptscriptstyle{\rm HMS}}}

\section{Introduction}\label{sec1}

The theory of General Relativity (GR), introduced by Albert Einstein in 1915 \cite{Einstein:1916vd}, stands as one of the most profound achievements in modern physics and is widely regarded as the most successful theory of gravity due to its remarkable agreement with experimental observations and predictions. GR fundamentally redefined our understanding of gravity, describing it not as a force but as the curvature of spacetime induced by mass and energy. This revolutionary framework has provided the foundation for numerous groundbreaking discoveries.
Among its early triumphs was the explanation of the anomalous precession of Mercury’s perihelion, an observation that had puzzled astronomers for decades and was resolved elegantly within the GR framework \cite{Will:2018mcj}. Another significant success was the prediction and subsequent observation of the bending of light by massive objects, a phenomenon confirmed during the solar eclipse of 1919 \cite{Will:2014zpa}. These early validations of GR solidified its status as a robust theory of gravitation.
In the modern era, GR has continued to prove its predictive power. The direct detection of gravitational waves by the LIGO and Virgo collaborations in 2015 \cite{LIGOScientific:2016aoc, LIGOScientific:2017ync} provided an unprecedented confirmation of Einstein’s theory, showing its ability to describe dynamic and extreme astrophysical events such as black hole mergers. Similarly, the Event Horizon Telescope (EHT) collaboration captured horizon-scale images of the supermassive black holes at the centers of the M87 galaxy and the Milky Way \cite{EventHorizonTelescope:2019dse, EventHorizonTelescope:2022wkp}, offering visual evidence of spacetime curvature near these exotic objects.

Despite its remarkable successes, GR is not without significant challenges. Central to these is its prediction of singularities, regions where the curvature of spacetime becomes infinite, and the equations of GR break down, losing their physical applicability. Singularities arise in critical scenarios, such as the Big Bang, which marks the initial state of the universe, and the cores of black holes. The latter, first described by Karl Schwarzschild in 1916 \cite{Schwarzschild}, are modeled as objects characterized solely by their mass, enclosed by an event horizon, a boundary beyond which not even light can escape due to the immense gravitational curvature. However, the Schwarzschild solution, like other classical black hole solutions in GR, is plagued by a central singularity where the laws of physics break down.
Resolving these singularities remains one of the most pressing challenges in modern gravitational physics. A complete theory of gravity must not only describe the dynamics of spacetime but also address these pathologies, ensuring that physical laws remain valid under all conditions. This challenge has driven substantial efforts to extend or modify GR, aiming to create a more comprehensive framework that eliminates singularities and deepens our understanding of gravity’s fundamental nature.
One avenue for advancing this understanding involves coupling Einstein’s equations with electromagnetism. This approach enables the exploration of how spacetime interacts with electromagnetic fields, revealing phenomena such as charged black holes and the unique properties they exhibit. By integrating electromagnetism into the equations of GR, we gain access to a rich variety of solutions, providing new insights into the interplay between gravitational and electromagnetic interactions in extreme astrophysical and cosmological settings.

In fact, an innovative approach to addressing the singularities inherent in Maxwell’s classical theory of electrodynamics was proposed by Born and Infeld in 1934 \cite{Born:1933pep,Born:1934gh}. They extended Maxwell’s framework to eliminate the singularities associated with a point charge and the infinite self-energy of such charges. This groundbreaking work led to the formulation of Nonlinear Electrodynamics (NLED), a theory that modifies the behavior of electromagnetic fields at high intensities. The Born-Infeld model set a foundation for further advancements, including significant contributions inspired by quantum field theory, such as the seminal Euler-Heisenberg theory \cite{Heisenberg:1936nmg}. Plebanski further expanded the theoretical framework of NLED, offering a broader generalization of its principles \cite{Plebanski}. Subsequent developments, such as those by Kruglov and others, have continued to refine and extend NLED, exploring its implications across various contexts \cite{Kruglov:2014hpa,Kruglov:2014iwa,Kruglov:2014iqa,Bandos:2020jsw}.
In 1968, a landmark advancement in black hole physics was introduced by Bardeen, who proposed the first regular black hole solution, one that avoids the central singularity traditionally associated with such objects \cite{Bardeen}. This pioneering concept laid the groundwork for a new class of solutions in gravitational theory. Decades later, the work of Ayon-Beato and García demonstrated that Bardeen’s solution could be derived as an exact solution of GR when the black hole's matter source was described using NLED \cite{Ayon-Beato:2000mjt}. In their interpretation, the regularization parameter in Bardeen’s model was attributed to the presence of a magnetic charge. Interestingly, subsequent work reinterpreted Bardeen’s solution as an NLED-driven GR solution, but with an electric charge as the source instead of a magnetic one \cite{Rodrigues:2018bdc}.

Inspired by these groundbreaking contributions, numerous models of regular black holes have been proposed that incorporate NLED as the matter source in GR. Notable examples include the models developed by Bronnikov \cite{Bronnikov:2000vy}, Dymnikova \cite{Dymnikova:2004zc}, Balart and Vagenas \cite{Balart:2014cga}, and Culetu \cite{Culetu:2014lca}. These models have expanded the landscape of regular black hole solutions, offering new insights into their physical properties and implications. Simultaneously, the thermodynamic behavior of regular black holes has attracted significant attention, with studies examining their entropy, temperature, and stability under various conditions \cite{Breton:2004qa,Myung:2007xd,Ma:2015gpa,Kruglov:2016ymq,Fan:2016hvf}.
For a broader perspective on the applications and implications of regular black hole models within the framework of NLED and GR, we refer the reader to the  works Balart, Kruglov, and Gullu, which delve into diverse scenarios and provide a comprehensive understanding of the subject \cite{Balart:2014jia,Kruglov:2016ezw,Kruglov:2017fck,Gullu:2020ant}. These studies collectively highlight the relevance and versatility of NLED in addressing longstanding challenges in theoretical physics, particularly in the quest to resolve singularities in both gravitational and electromagnetic contexts.
In the study by \cite{Bokulic:2022cyk}, the authors investigate the inherent limitations of the NLED solutions when attempting to regularize black hole singularities. They rigorously demonstrate that several classes of Lagrangians—including more general cases with joint dependence on the electromagnetic invariants $F$ and $G$—prevent the complete elimination of curvature singularities in static and spherically symmetric solutions with both electric and magnetic charges. These results reinforce the intrinsic difficulties in using nonlinear electromagnetic fields to avoid the formation of singularities within classical models, clearly highlighting the limitations of such approaches in the search for regular black hole solutions based on classical electromagnetic field theories.
Furthermore, in addition to these studies and applications, there is a comprehensive review of regular black holes, in which the author discusses key theoretical aspects and recent developments in the field, offering a consolidated and up-to-date perspective on the subject \cite{Lan:2023cvz}.

Another notable development emerged in 1966, when Schwinger introduced the concept of a particle possessing both magnetic and electric charges \cite{Schwinger:1966nj}. This particle, now widely known as the dyon or dyonic solution, has since become a key element in theoretical physics. Dyonic solutions have been applied across a wide range of contexts, for instance, they have been employed in analyzing the properties of rotating black holes within the Tomimatsu-Sato-Yamazaki spacetime framework in the context of dyons \cite{Kasuya:1981ef}. They have also been used to study radiating massive objects \cite{Chamorro:1995bv}, and their applications extend to string theory \cite{Garfinkle:1990qj,Sen:1992fr,Cheng:1993wp,Mignemi:1993ce,Lowe:1994gt,Jatkar:1995ut,Cvetic:1995bj,Tseytlin:1996as} and supergravity theories \cite{Chamseddine:2000bk,Chow:2013gba,Lu:2013ura,Guarino:2015jca,Benini:2016rke,Meessen:2017rwm}. For readers interested in a deeper exploration of dyonic applications, we recommend consulting the extensive works available on the subject \cite{Shapere:1991ta,Poletti:1995yq,Brihaye:1998cm,LopesCardoso:2004law,Hartnoll:2007ai,Hartnoll:2007ih,Albash:2008eh,Chen:2008hk,Caldarelli:2008ze,Chen:2010yu,Dutta:2013dca,Cai:2014oca,Li:2016nll,Hendi:2016uni,Bronnikov:2017xrt,Bambi:2023try}.

Indeed, in 2018, Simpson and Visser proposed a novel approach to address singularities in GR solutions through a framework now known as the Black Bounce (BB) model \cite{Simpson:2018tsi}. Motivated by regularizing parameters that eliminate the central singularity at 
$r=0$, their proposal introduces a ``bounce'' structure mediated by one or more parameters related to a fundamental length scale. This regularization ensures that the spherical area radius remains non-zero, preventing the formation of a singularity and allowing for a smooth transition in the radial coordinate.
The BB framework accommodates a variety of configurations, making it a powerful tool in gravitational physics. 
Among the notable solutions are regular black holes with symmetric horizons and a central bounce, one-way traversable wormholes, and two-way traversable wormholes. These configurations provide rich opportunities to explore spacetime geometries free from singularities, while also uncovering new physical phenomena.
The BB framework has inspired the development of numerous extensions, alternative solutions, and analytical models, leading to significant advancements in the field. Research has focused on aspects such as the regularity of spacetime, causal structures, and energy conditions \cite{Lobo:2020ffi}; gravitational lensing effects \cite{Nascimento:2020ime, Tsukamoto:2020bjm, Cheng:2021hoc, Tsukamoto:2021caq, Zhang:2022nnj}; and potential observational signatures \cite{Guerrero:2021ues, Jafarzade:2021umv, Jafarzade:2020ova, Jafarzade:2020ilt, Yang:2021cvh, Bambhaniya:2021ugr, Ou:2021efv, Guo:2021wid, Wu:2022eiv, Tsukamoto:2022vkt}. Furthermore, generalizations of the BB model to rotating configurations have also been explored, extending its applicability to more realistic astrophysical scenarios \cite{Mazza:2021rgq, Xu:2021lff}.

Black bounce (BB) configurations are typically obtained using an inverse method, wherein a predefined line element exhibiting bounce-like characteristics is first specified. The corresponding field equations are then solved in reverse to identify the matter sources that give rise to the given spacetime geometry. In many cases, these material sources are described by a combination of NLED and scalar fields \cite{Bronnikov:2022bud, Canate:2022gpy, Rodrigues2023, Pereira:2023lck}. Historically, most BB solutions have used only a magnetic charge as the primary component of the Maxwell-Faraday tensor.
More recently, however, BB solutions have been constructed using material sources with a purely electric charge embedded within the NLED structure \cite{Alencar:2024yvh}. This development opens the door to studying configurations that can exhibit both electric and magnetic charges, significantly broadening the scope of BB solutions and their applications. These developments highlight the ongoing refinement of the BB model and its potential to reveal novel insights into gravitational theory.

In this work, our primary objective is to investigate dyonic solutions within the framework of BB geometries in GR, incorporating both magnetic and electric charges in the structure of NLED, following the approach described by Bronnikov \cite{Bronnikov:2000vy}. Specifically, we focus on static and spherically symmetric configurations, coupling NLED and a scalar field as matter sources in GR.
Our analysis involves the development of BB solutions for various metric functions, where we numerically examine the existence and nature of horizons, assess the regularity of the solutions through the behavior of the Kretschmann scalar, and determine the material sources responsible for shaping the geometric structure of these configurations. Through this approach, we aim to provide a comprehensive understanding of dyonic BB solutions and their physical implications within this extended framework.

Recently, other formulations have been proposed that aim to avoid the singularity of black holes. In this context, the paper \cite{Capozziello:2024ucm} investigated a Schwarzschild-type metric that exhibits a change in signature when crossing the event horizon, giving rise to the so-called Lorentzian-Euclidean black hole. The resulting geometry is regularized using Hadamard's partie-finie technique, which allows us to show that this metric is a solution of Einstein's equations in vacuum. In the context of this construction, the concept of timelessness was introduced, which is understood as a dynamical mechanism responsible for the transition from one regime with real time variables to another with imaginary time. As a result, this mechanism eliminates the presence of the central singularity characteristic of classical black holes.

The analysis of the stability of regular black holes in NLED theories has revealed significant challenges related to the emergence of instabilities near the central region. In the study by Felice and Tsujikawa \cite{DeFelice:2024seu}, the authors investigate the stability of nonsingular black holes in an NLED framework coupled to general relativity. They demonstrate that although these solutions are free from curvature singularities, they exhibit an angular Laplacian instability near the center, associated with a negative squared sound speed in the angular direction. This instability leads to the exponential growth of vector perturbations, ultimately compromising the stability of the regular background spacetime.
Subsequently, in \cite{DeFelice:2024ops}, the same authors extend this analysis by considering a more general theory in which NLED is coupled to a scalar field $\phi$, with the Lagrangian expressed as ${\cal L}(F, \phi, X)$, where $X$ represents the kinetic term of the scalar field. In this work, they also examine the stability of static and spherically symmetric black hole solutions, finding that achieving configurations that are both nonsingular and linearly stable is highly challenging. Moreover, consistent with their previous results, they do not find solutions that simultaneously support both electric and magnetic charges (dyons), further reinforcing the inherent difficulties in constructing stable regular black hole models within this class of classical theories.

Despite these recent advances, the challenge of constructing dyonic regular solutions for black holes that are simultaneously free of singularities and dynamically stable remains largely unsolved. In particular, the limitations of NLED-based models as well as the conceptual complexity associated with signature change approaches motivate the search for alternative frameworks.

In this context, the present paper introduces a novel black bounce solution in the dyonic sector, which simultaneously incorporates electric and magnetic charges coupled to a scalar field with potential. This is the first solution of this kind reported in the literature, derived from the coupling between a linear Lagrangian for the electromagnetic sector and a scalar field that acts as the matter source sustaining the regular spacetime. This construction introduces an effective regularization mechanism that removes the central singularity, replacing it with a smooth geometric transition at the core. Furthermore, the study demonstrates that, unlike the persistent challenges in approaches based on nonlinear electrodynamics (NLED)—where obtaining regular and stable dyonic solutions is particularly difficult—the solution proposed here is free from central singularities. This result represents a significant contribution to the development of regular black hole models within the framework of General Relativity.

This work is organized as follows: In Sec. \ref{sec2}, we provide a brief overview of the field equations for GR coupled with NLED and scalar fields, explicitly incorporating both magnetic and electric charges. In Secs. \ref{BB_SV} to \ref{solV}, we analyze and generalize various BB geometries, presenting several static and spherically symmetric solutions derived within this framework. Finally, in Sec. \ref{sec:concl}, we conclude by summarizing our results and discussing their implications.

\section{Field equations coupled to non-Linear slectrodynamics and scalar field}\label{sec2}

\subsection{Action and field equations}

We begin our analysis with the following action:
\begin{align}
S	=\int\sqrt{-g}d^{4}x \bigg[&R-2\kappa^{2}\Big({\cal L}(\varphi)-{\cal L}_{\textrm{NLED}}(F)\Big)\bigg],
\label{action}
\end{align}
with $\kappa^2=8\pi$ and
\begin{equation}
    {\cal L}(\varphi)=\epsilon \, \partial^{\mu}\varphi\partial_{\mu}\varphi-V\left(\varphi\right),
\end{equation}
where $\epsilon=+1$ corresponds to a regular scalar field. while $\epsilon=-1$ represents a phantom scalar field. The field $\varphi$ denotes the scalar field, while $V(\varphi)$ is the associated scalar potential. Additionally, ${\cal L}_{\rm NLED}(F)$ is the Lagrangian density that describes a NLED, which depends on the electromagnetic field scalar
\begin{equation}
F=\frac{1}{4}F^{\mu\nu}F_{\mu\nu},    \label{F}
\end{equation}
where the  Maxwell-Faraday antisymmetric tensor is defined by
\begin{equation}
   F_{\mu \nu} = \partial_\mu A_\nu -\partial_\nu A_\mu, \label{MaxFar}
\end{equation}
 and ${A_\mu}$ is the magnetic vector potential.

Presented below are the equations of motion arising from the action \eqref{action}:
\begin{align}
\nabla_\mu ({\cal L}_F F^{\mu\nu})&=0,\label{sol2}\\
\nabla_\mu \,{}^\star\! F^{\mu\nu}&=0,\label{sol2b}\\
     2\epsilon\nabla_\mu\nabla^\mu \varphi&=-\frac{dV(\varphi)}{d\varphi}\,,\label{sol3}
\end{align}
where ${\cal L}_F=\partial {\cal L} _{\rm NLED}(F)/\partial F$, ${}^{\star}\!F^{\mu\nu}=\frac{1}{2\sqrt{-g}}\varepsilon^{\mu\nu\alpha\beta}\!F_{\alpha\beta}$ and  the asterisk denotes the Hodge dual ($^\star$).
\par
The gravitational field equations are obtained by varying the action \eqref{action} with respect to the metric tensor, which yield:
\begin{equation}
G_{\phantom{\mu}\nu}^{\mu}=R_{\phantom{\mu}\nu}^{\mu}-\frac{1}{2}\delta{}_{\phantom{\mu}\nu}^{\mu}R=\kappa^{2}\left(\overset{F}{T}{}_{\phantom{\mu}\nu}^{\mu}+\overset{\varphi}{T}{}_{\phantom{\mu}\nu}^{\mu}\right).\label{EqM}
\end{equation}
Thus, the explicit form of the NLED energy-momentum tensor is expressed as follows
\begin{align}
    \overset{F}{T}{}_{\phantom{\mu}\nu}^{\mu}=&\frac{2}{\kappa^{2}}\Big(\delta_{\phantom{\mu}\nu}^{\mu}{\cal L}_{{\rm NLED}}(F)-{\cal L}_{F}F^{\mu\alpha}F_{\nu\alpha}\Big),
\end{align}
and the energy-momentum tensor for the scalar field matter part is defined as
\begin{equation}
    \overset{\varphi}{T}{}_{\phantom{\mu}\nu}^{\mu}=\epsilon\,\Big(2\,\partial^{\mu}\varphi\partial_{\nu}\varphi-\delta_{\phantom{\mu}\nu}^{\mu}\partial^{\sigma}\varphi\partial_{\sigma}\varphi\Big)+\delta_{\phantom{\mu}\nu}^{\mu}V(\varphi).
\end{equation}

In this work, we consider the following static and spherically symmetric metric to obtain our solutions:
\begin{equation}
ds^2=A(r)dt^2-\frac{1}{A(r)}dr^2-\Sigma^2(r) \, d\Omega^2,\label{m}
\end{equation}
where $A(r)$ and $\Sigma(r)$ are functions of the radial coordinate $r$, and the two-sphere line element is defined as $d\Omega^2\equiv d\theta^{2}+\sin^{2}\left(\theta\right)d\phi^{2}$. 
Note that the metric function $\Sigma(r)$ allows us to express the metric in its most general form for spherical and static symmetry, as given in Eq.\,\eqref{m}. In the sections that follow, we present the corresponding metrics for the solutions describing black holes and regular black holes.

Furthermore, we consider solutions described by the magnetic charge, $q_m$, and the electric charge $q_e$. The components of the $F_{\mu\nu}$  tensor are described by:
\begin{align}
    F_{01}=-F_{10}&=\frac{q_{e}}{\Sigma^{4}(r){\cal L}_F(r)}  \,,\\
    F_{23}=-F_{32}&=q_m \sin\theta \,,
\end{align}
and the electromagnetic scalar takes the form
\begin{equation}
    F=\frac{q_{m}^{2}{\cal L}_{F}^{2}(r)-q_{e}^{2}}{2{\cal L}_{F}^{2}(r)\Sigma(r)^{4}}.\label{F2}
\end{equation}

\subsection{Strategy for solving the field equations}

To assess the consistency of our solutions, we will also use the following relationship
\begin{equation}
    {\cal L}_F=\frac{\partial {\cal L} _{\rm NLED}}{\partial r} \bigg(\frac{\partial F}{\partial r}\bigg)^{-1}.\label{RC}
\end{equation}

The components derived from the equations of motion, as described by Eq.\,\eqref{EqM}, after substituting Eq.\,\eqref{m}, are given by:
\begin{eqnarray}
	&&-\frac{\Sigma(r)\left(A'(r)\Sigma'(r)+2A(r)\Sigma''(r)\right)+A(r)\Sigma'(r)^{2}-1}{\Sigma(r)^{2}}
		\nonumber \\
	&& \qquad  =\kappa^{2}\Big(\epsilon A(r)\varphi'(r)^{2}+V(r)\Big)
		\nonumber \\
	&& \qquad \qquad +2\left({\cal L}_{\text{NLED}}(r)+\frac{q_{e}^{2}}{{\cal L}_{F}(r)\Sigma(r)^{4}}\right)\,,
		\label{EqF00}
\end{eqnarray}
\begin{eqnarray}
	&&-\frac{\Sigma(r)A'(r)\Sigma'(r)+A(r)\Sigma'(r)^{2}-1}{\Sigma(r)^{2}}
		\nonumber \\
	&& \qquad  =\kappa^{2}\Big(V(r)-\epsilon A(r)\varphi'(r)^{2}\Big)
		\nonumber \\
	&& \qquad \qquad +2\left({\cal L}_{\text{NLED}}(r)+\frac{q_{e}^{2}}{{\cal L}_{F}(r)\Sigma(r)^{4}}\right)\,,
	\label{EqF11}
\end{eqnarray}
\begin{eqnarray}
	&&-\frac{\Sigma(r)A''(r)+2A'(r)\Sigma'(r)+2A(r)\Sigma''(r)}{2\Sigma(r)}
		\nonumber \\
	&& \qquad =\kappa^{2}\Big(\epsilon A(r)\varphi'(r)^{2}+V(r)\Big)
		\nonumber \\
	&& \qquad \qquad +2\left({\cal L}_{\text{NLED}}(r)-\frac{q_{m}^{2} {\cal L}_{F}(r)}{\Sigma(r)^{4}}\right).\label{EqF22}
\end{eqnarray}

We can solve the set of equations (\ref{EqF00})–(\ref{EqF22}) using two different approaches.  
In the first approach, we derive the general expressions for ${\cal L}_{\rm NLED}(r)$ and ${\cal L}_{F}(r)$ from the equations of motion (\ref{EqF00}) and (\ref{EqF22}). This process ensures that only the equation of motion \eqref{EqF11} remains, which we then use to determine the scalar field $\varphi(r)$.  
Following this strategy, we obtain the following quantities directly from the equations of motion \eqref{EqF00} and \eqref{EqF22}:
\begin{widetext}
\begin{eqnarray}
&&{\cal L}_{\rm NLED}(r) =
\frac{1}{8\Sigma(r)^{4}}
\Bigg\{-\Sigma(r)^{4}\left[A''(r)+4\kappa^{2}\Big(\epsilon A(r)\varphi'(r)^{2}+V(r)\Big)\right]-2\Sigma(r)^{3}\left(2A'(r)\Sigma'(r)+3A(r)\Sigma''(r)\right)
\nonumber\\
&& \qquad \qquad -2\Sigma(r)^{2}\left(A(r)\Sigma'(r)^{2}-1\right)
+
\sqrt{\Sigma(r)^{4}\Big[\Sigma(r)^{2}A''(r)-2A(r)\Big(\Sigma(r)\Sigma''(r)+\Sigma'(r)^{2}\Big)+2\Big]^{2}-64 q_{e}^{2}q_{m}^{2}}\Bigg\},	\label{L_BB} 
 \end{eqnarray}
and
 \begin{eqnarray}
&& {\cal L}_F(r) = 
\frac{1}{8q_{m}^{2}}
\Bigg\{
\Sigma(r)^{4}A''(r)-2A(r)\Sigma(r)^{3}\Sigma''(r)-2\Sigma(r)^{2}\left(A(r)\Sigma'(r)^{2}-1\right)
\nonumber\\	&&
+\sqrt{\Sigma(r)^{4}\Big[\Sigma(r)^{2}A''(r)-2A(r)\Big(\Sigma(r)\Sigma''(r)+\Sigma'(r)^{2}\Big)+2\Big]^{2}-64q_{e}^{2}q_{m}^{2}}\; \Bigg\}	.\label{LF_BB}
\end{eqnarray}

\end{widetext}

If we now insert Eqs.\;\eqref{L_BB} and \eqref{LF_BB} into the equation of motion \eqref{EqF11}, we get
\begin{equation}
   \frac{2 A(r) \left(\Sigma ''(r)+\kappa ^2 \epsilon  \Sigma (r) \varphi '(r)^2\right)}{\Sigma (r)}=0.\label{exp11}
\end{equation}
Thus, as stated above, Eq.~\eqref{exp11} can be solved to determine the general form of the scalar field $\varphi(r)$, whose explicit expression is presented below
\begin{equation}
    \varphi(r)=\int\frac{\sqrt{\Sigma''(r)}}{\kappa\sqrt{\left(-\epsilon\right)\Sigma(r)}}\,dr.\label{varphi}
\end{equation}  

With this scalar field, the potential $V(r)$ can now be determined after solving Eq.~\eqref{sol3}, which yields the following:
\begin{align}
    V(r)=\int&\frac{1}{\Sigma(r)^{2}}\Big[\Sigma''(r)\Big(2\Sigma(r)A'(r)+3A(r)\Sigma'(r)\Big)
    \nonumber\\&
    +A(r)\Sigma(r)\Sigma^{'''}(r)\Big]\,dr.\label{V}
\end{align}

The second strategy follows an analogous approach to the first. However, in this case, we use the equations of motion \eqref{EqF11} and \eqref{EqF22} to determine the general forms of ${\cal L}_{\rm NLED}(r)$ and ${\cal L}_F(r)$. These expressions can then be substituted into the equation of motion \eqref{EqF00} to obtain the same scalar field $\varphi(r)$ given by \eqref{varphi}, and consequently, the same potential as in \eqref{V}.  
Thus, we conclude that the solutions obtained using the second strategy for $A(r)$, $\Sigma(r)$, ${\cal L}_{\rm NLED}(r)$, ${\cal L}_{F}(r)$, $\varphi(r)$, and $V(r)$ are identical to those obtained with the first approach.

To analyze the properties of the metric functions, which will be presented later, and to determine the presence of horizons, we impose the following condition:
\begin{equation}
     A(r_{H})=0, \label{rH}
\end{equation}
where the radius $r_{H}$ denotes the radial coordinate of the horizon. Moreover, a second condition allows one to compute the number of horizons of a given metric function
\begin{equation}
    \frac{d A(r_{H})}{dr}\bigg|_{r=r_H}=0.
    \label{der_a}
\end{equation}

These two relations allow us to identify degenerate horizons and subsequently determine the critical values of the parameters for each model (i.e. mass $M$, electric charge $q_e$, or magnetic charge $q_m$), which will be obtained in the following sections. To achieve this, we will numerically solve Eqs. \eqref{rH} and \eqref{der_a} simultaneously to obtain the critical value of one of the geometry's free parameters, while fixing the values of the other two. This is done for all three possible free parameters. In this way, we will analyze additional solutions based on the chosen parameter values, thus exploring the available range of geometric structures.

Additionally, we will calculate the Komar mass for our models~\cite{Komar:1958wp}. This quantity is essential for determining the mass of a stationary, asymptotically flat spacetime and is directly related to the Killing field. The Komar mass is determined by the following surface integral
\begin{equation}
    \bar{M}_K=-\frac{1}{8\pi}\int_\infty \star \, dk,\label{m_Komar}
\end{equation}
where $k =  g_{tt} dt$ is the covector associated with the Killing field asymptotically similar to the time field $\partial_t$. 

For a metric that depends only on the radial coordinate $r$, as is the case discussed in this manuscript and illustrated by the metric \eqref{m}, we can simplify the integral \eqref{m_Komar} as follows
\begin{equation}
M_{K} = \lim_{r \rightarrow \infty} \frac{1}{2} r^{2} \partial_{r} A(r). \label{m_Komar2}
\end{equation}

We shall use Eq. \eqref{m_Komar2} to calculate the Komar mass in our models.

\subsection{Dyonic Reissner-Nordström type solution}

As an example, and a first solution, we assume that the contribution of the scalar field is zero, i.e. $V(r)=0$ and $\varphi (r)=0$. We also assume that $\Sigma(r)=r$ and that the Lagrangian and its derivative are
\begin{align}
    {\cal L}_{\rm NLED}(r) =F,\\
    {\cal L}_{F}(r) =1.
\end{align}

With these assumptions, we obtain the following equations of motion:
\begin{align}
    -\frac{r A'(r)+A(r)-1}{r^2}=\frac{q_e^2+q_m^2}{r^4},\\
    \frac{A''(r)}{2}=\frac{r^3 \left(-A'(r)\right)+q_e^2+q_m^2}{r^4}\,,
\end{align}
which provide us with the following solution
\begin{equation}
    A(r)=1-\frac{2 M}{r}+\frac{q_e^2+q_m^2}{r^2}.\label{aRN}
\end{equation}
This metric function satisfies all equations of motion and is known as the dyonic Reissner-Nordström solution. If we consider the magnetic charge to be zero ($q_m=0$), we obtain the Reissner-Nordström metric function.

The Komar mass for this model, given by Eq. \eqref{m_Komar2}, is:
\begin{equation}
M_K = \lim_{r \rightarrow \infty} \left(M - \frac{ \left(q_{e}^{2} + q_{m}^{2}\right)}{r}\right) = M. \label{MK_RN}
\end{equation}
Thus, the parameter $M$, described by Eq. \eqref{aRN}, is identical to the Komar mass given by Eq. \eqref{MK_RN}.

\section{Dyon solutions of the Simpson-Visser type}\label{BB_SV}

In this model, we generalize the Simpson-Visser model by incorporating both the magnetic charge $q_m$ and the electric charge $q_e$. However, due to the direct algebraic construction employed in deriving this solution, certain inconsistencies may arise. For instance, the Lagrangian $\mathcal{L}(r)$ and its derivative $\mathcal{L}_F(r)$ are not real over the entire range of the radial coordinate $r$. We will further discuss these results in the following sections.

\subsection{Dyon solutions of the Simpson-Visser type}

In this approach, we extend the metric function of the Simpson-Visser BB model to the dyonic framework to derive our solutions. In this process, we consider that
\begin{equation}
A(r)=1-\frac{2M}{\sqrt{q_{e}^{2}+q_{m}^{2}+r^{2}}}.\label{A_SV}
\end{equation}
Setting the electrical charge $q_e=0$ in Eq.\,\eqref{A_SV}, yields the Simpson-Visser solution.

With the purpose of generating a bounce, we can choose the canonical form of the area function described as follows,
\begin{equation}
\Sigma(r)=\sqrt{q_m^{2}+r^{2}},\label{Sig0}
\end{equation}
where the coordinate $r$ ranges over $(-\infty, +\infty)$, while the function $\Sigma(r)$, which defines the areal radius of the two-spheres, is given by $S = 4\pi \Sigma^2(r)$ and assumes a minimum value of $\Sigma(r) = q_m$ at $r = 0$.

As we develop dyon solutions in this manuscript we will conveniently consider an appropriate generalization of the function described by Eq. \eqref{Sig0}. Namely, we will adopt the following form
\begin{equation}
\Sigma(r)=\sqrt{q_{e}^{2}+q_{m}^{2}+r^{2}}.\label{Sig1}
\end{equation}
With this definition, the minimum value of Eq. \eqref{Sig1} occurs when $\Sigma(0) = \sqrt{q_{e}^{2} + q_{m}^{2}}$, and we recover the Simpson-Visser area function, when we set $q_e=0$ in Eq.\,\eqref{Sig1}.

\begin{figure*}[!th]
\includegraphics[scale=0.56]{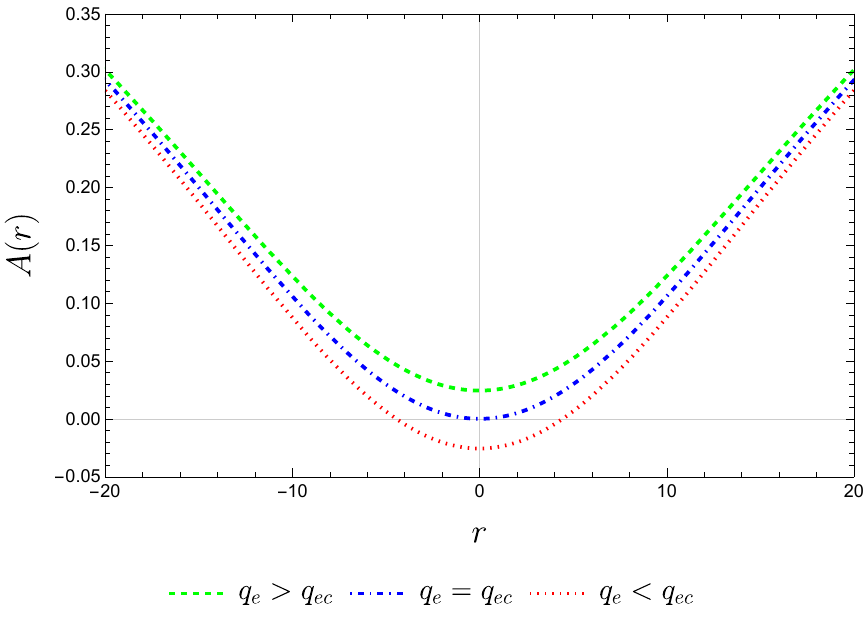}
\hspace{0.75cm}
\includegraphics[scale=0.56]{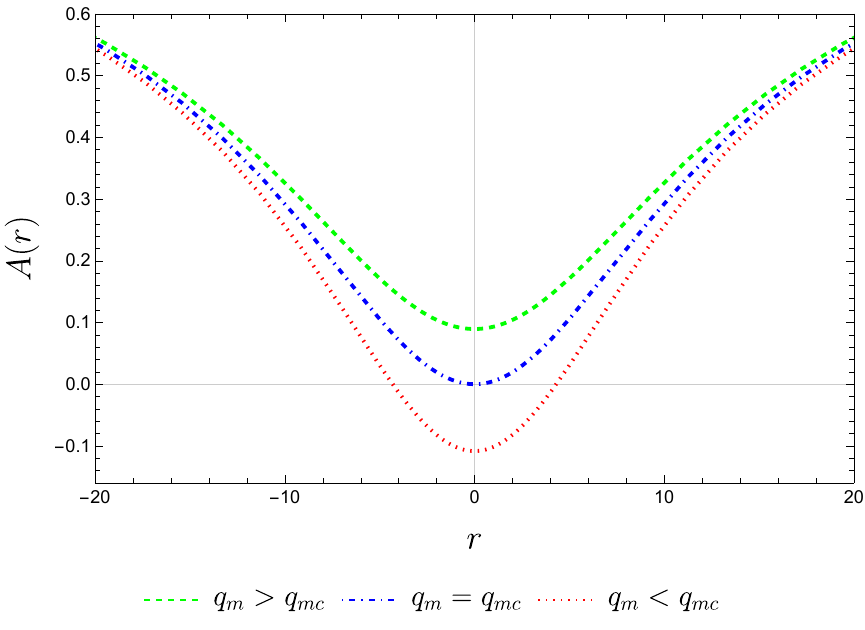}
\caption{The  metric function $A(r) $, given by Eq.~\eqref{A_SV}. (i) The left plot depicts three electric charge scenarios, where we have used the values of the constants as follows $\{M=10.0$ and $q_m=0.2$. (ii) The right plot depicts three magnetic charge scenarios, with the following values $M=5.0$ and $q_e=2.0$.} 
\label{aqc_SV}
\end{figure*}

Now, we simultaneously solve Eqs. \eqref{rH} and \eqref{der_a} to determine the critical electric charge, which is $q_{ec}=19.999$ when we assign the constant values $M=10.0$ and $q_m=0.2$. In the left plot of Fig. \ref{aqc_SV}, we illustrate the behavior of the metric function \eqref{A_SV} with respect to the radial coordinate $r$ for three different scenarios of the electric charge: $q_e > q_{ec}$, $q_e = q_{ec}$ and $q_e < q_{ec}$.
  
\begin{figure}[!th]
\includegraphics[scale=0.56]{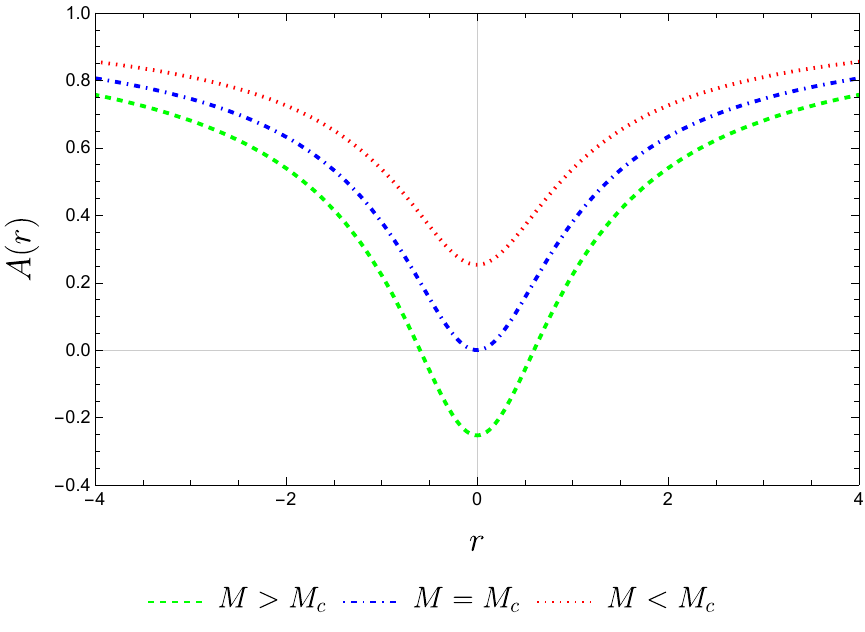}
\caption{The  metric function $A(r) $, given by Eq.~\eqref{A_SV} for three mass scenarios. We have used the values of the constants as follows $\{q_e=0.25$. and $q_m=0.75\}$ } 
\label{amassc_SV}
\end{figure}

Similarly, by simultaneously solving Eqs.\,\eqref{rH} and \eqref{der_a}, we find the critical magnetic charge $q_{mc}=9.797$ when we choose the constant values $M=5.0$ and $q_e=2.0$. We illustrate the behavior of the metric function \eqref{A_SV} for this case in the right plot of Fig. \ref{aqc_SV} considering the scenarios of magnetic charge: $q_m > q_{mc}$, $q_m = q_{mc}$ and $q_m < q_{mc}$. We observe a similar behavior in all three scenarios, when varying the corresponding free parameter, as illustrated in Figs. \ref{aqc_SV} and \ref{amassc_SV}.

The critical mass $M_c$, determined by solving Eqs. (\ref{rH}) and (\ref{der_a}) simultaneously, is found to be $M_c = 0.395$ when we choose the constant values as $q_e=0.25$ and $q_m=0.75$. Fig.~\ref{amassc_SV} shows the behavior of the metric function \eqref{A_SV} in terms of the radial coordinate $r$, for three mass scenarios: $M > M_c$, $M = M_c$, and $M < M_c$. 

Using Eq. \eqref{m_Komar2} we computed the Komar mass, for the metric function \eqref{A_SV}. In this case, we obtained:
\begin{equation}
M_K = \lim_{r \rightarrow \infty} \frac{Mr^{3}}{\left(q_{e}^{2} + q_{m}^{2} + r^{2}\right)^{3/2}} = M. \label{MK_SV}
\end{equation}
Thus, the parameter $M$, described by Eq.\,\eqref{A_SV}, is identical to the Komar mass given by Eq.\,\eqref{MK_SV}.

\subsection{Kretschmann scalar}

\begin{figure}[!th]
\includegraphics[scale=0.56]{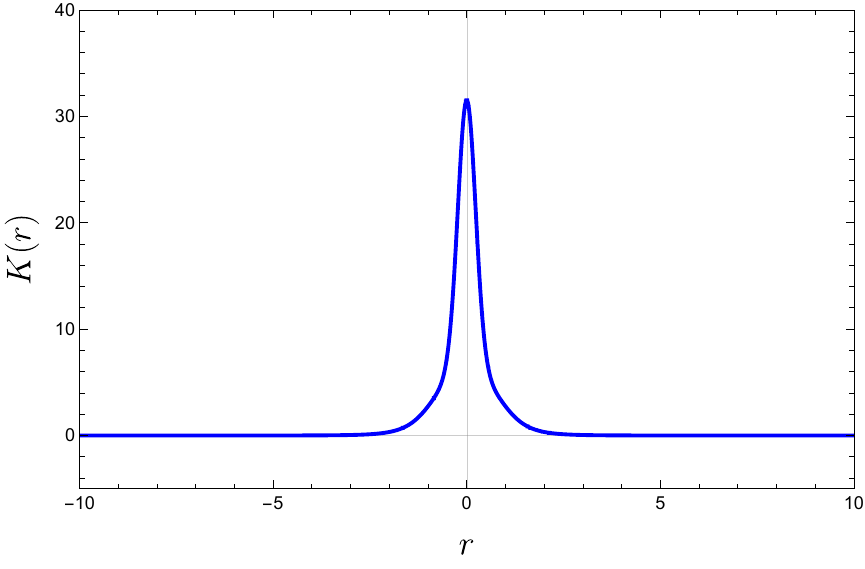}
\caption{The scalar of Kretschmann $K(r) $, given by Eq.~\eqref{K_SV}. We have used the values of the constants as follows $\{M=1.0,\,q_e=0.5,\,q_m=0.75\}$.  } 
\label{fig_K_SV}
\end{figure}

On the other hand, we can study the regularity properties of our model by computing the  Kretschmann scalar, which for the present case is given by
\begin{eqnarray}
&&K(r)= \frac{4}{\left(q_{e}^{2}+q_{m}^{2}+r^{2}\right)^{5}} \Big\{
3\left(q_{e}^{2}+q_{m}^{2}\right)^{2}\left(q_{e}^{2}+q_{m}^{2}+r^{2}\right)
\nonumber\\
&&
\quad -8M\left(q_{e}^{2}+q_{m}^{2}\right)\left(q_{e}^{2}+q_{m}^{2}-r^{2}\right)\sqrt{q_{e}^{2}+q_{m}^{2}+r^{2}}
\nonumber\\
&&
\quad +3M^{2}\left[-4r^{2}\left(q_{e}^{2}+q_{m}^{2}\right)+3\left(q_{e}^{2}+q_{m}^{2}\right)^{2}+4r^{4}\right]\bigg\}.
\label{K_SV}
\end{eqnarray}
Note that the Kretschmann scalar is regular for small values of $r$, i.e. in the limit $r \to 0$
\begin{equation}
\lim_{r\rightarrow0}K\left(r\right)=\frac{4\left(9M^{2}-8M\sqrt{q_{e}^{2}+q_{m}^{2}}+3q_{e}^{2}+3q_{m}^{2}\right)}{\left(q_{e}^{2}+q_{m}^{2}\right)^{3}}\,.\label{K_SVb}
\end{equation}
We observe that the Kretschmann scalar is asymptotically flat for very large values of $r$, i.e. in the limit $r \to \infty$. We illustrate the behavior of the Kretschmann scalar for this model in Fig. \ref{fig_K_SV}.

\subsection{Matter Lagrangian and Potential reconstruction}

By considering the model described by Eq.~\eqref{Sig1} and substituting it into the scalar field equation \eqref{varphi}, we obtain $\varphi(r)$, which will be used in the subsequent models discussed below:
\begin{equation}
\varphi(r)=\frac{\tan^{-1}\left(\frac{r}{\sqrt{q_{e}^{2}+q_{m}^{2}}}\right)}{\sqrt{\kappa^{2}(-\epsilon)}}.\label{field_camp}
\end{equation}
Note that if we consider zero electric charge in the scalar field above, Eq.\,\eqref{field_camp}, we recover the scalar field obtained in the work of \cite{Rodrigues2023}. 

By substituting the model \eqref{Sig1} and the metric function \eqref{A_SV} into Eq.~\eqref{V}, we obtain the expression for the potential as a function of the radial coordinate:
\begin{align}
    V(r)=&\frac{4M\left(q_{e}^{2}+q_{m}^{2}\right)}{5\kappa^{2}\left(q_{e}^{2}+q_{m}^{2}+r^{2}\right)^{5/2}}
	.\label{V_SV}
\end{align}

With the quantities such as the function \eqref{Sig1}, the metric function \eqref{A_SV}, the scalar field \eqref{field_camp} and the potential \eqref{V_SV}, we can explicitly express the Lagrangian ${\cal L}_{\textrm{NLED}}$ for this model, which is derived from the general Lagrangian given by Eq. \eqref{L_BB}, as
\begin{align}
{\cal L}_{\textrm{NLED}}(r) =&\, \frac{\sqrt{\frac{9M^{2}\left(q_{e}^{2}+q_{m}^{2}\right)^{2}}{q_{e}^{2}+q_{m}^{2}+r^{2}}-16q_{e}^{2}q_{m}^{2}}}{4\left(q_{e}^{2}+q_{m}^{2}+r^{2}\right)^{2}}	
\nonumber\\
&
-\frac{3M\left(q_{e}^{2}+q_{m}^{2}\right)}{20\left(q_{e}^{2}+q_{m}^{2}+r^{2}\right)^{5/2}}	,
\label{L_SV} 
\end{align}
and its derivative ${\cal L}_F$ as
\begin{align}
{\cal L}_F(r) =&\,\frac{\sqrt{\frac{9M^{2}\left(q_{e}^{2}+q_{m}^{2}\right)^{2}}{q_{e}^{2}+q_{m}^{2}+r^{2}}-16q_{e}^{2}q_{m}^{2}}+\frac{3M\left(q_{e}^{2}+q_{m}^{2}\right)}{\sqrt{q_{e}^{2}+q_{m}^{2}+r^{2}}}}{4q_{m}^{2}}.\label{LF_SV}
\end{align}

With the metric function \eqref{A_SV}, the function \eqref{Sig1}, the scalar field \eqref{field_camp}, the potential \eqref{V_SV}, the Lagrangian \eqref{L_SV}, and its derivative \eqref{LF_SV}, we now satisfy all equations\;\eqref{sol2}–\eqref{sol3} and \eqref{RC}, as well as all components of the equations of motion \eqref{EqF00}–\eqref{EqF22}.


Now the electromagnetic scalar \eqref{F} for this solution has the form of:
\begin{align}
&F(r)=q_m^2\Bigg[
\left(q_{e}^{2}+q_{m}^{2}\right)\left(\frac{9M^{2}\left(q_{e}^{2}+q_{m}^{2}\right)^{2}}{q_{e}^{2}+q_{m}^{2}+r^{2}}-16q_{e}^{2}q_{m}^{2}\right)^{\frac{1}{2}}
\nonumber\\&
\qquad \times3M\sqrt{q_{e}^{2}+q_{m}^{2}+r^{2}}-16q_{e}^{2}q_{m}^{2}\left(q_{e}^{2}+q_{m}^{2}+r^{2}\right)
\nonumber\\&
+9M^{2}\left(q_{e}^{2}+q_{m}^{2}\right)^{2}\Bigg]\Big/\left(q_{e}^{2}+q_{m}^{2}+r^{2}\right)^{2}
\Bigg[3M\left(q_{e}^{2}+q_{m}^{2}\right)
\nonumber\\&
+\sqrt{q_{e}^{2}+q_{m}^{2}+r^{2}}
\sqrt{\frac{9M^{2}\left(q_{e}^{2}+q_{m}^{2}\right)^{2}}{q_{e}^{2}+q_{m}^{2}+r^{2}}-16q_{e}^{2}q_{m}^{2}}\Bigg]^2
.\label{scalF_SV}
\end{align}
Note that we recover the electromagnetic scalar of the Simpson-Visser solution by taking $q_e=0$ in Eq.\;\eqref{scalF_SV}.
We highlight the behavior of the electromagnetic scalar \eqref{scalF_SV} in two distinct scenarios. 

\begin{figure*}[ht!]
\centering
\includegraphics[scale=0.56]{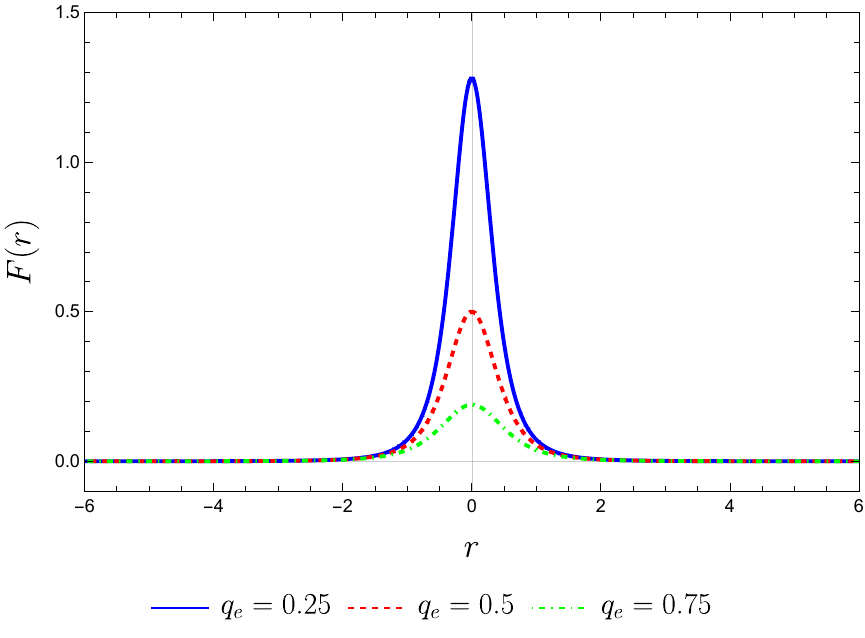}
\hspace{0.75cm}
\includegraphics[scale=0.56]{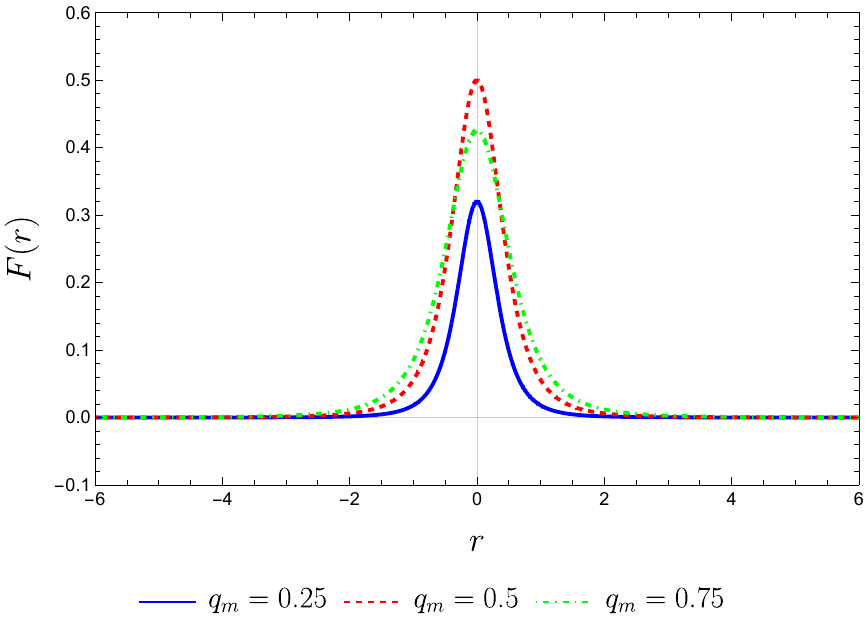}
\caption{The electromagnetic scalar $F$, given by Eq.~\eqref{scalF_SV}. (i) The left plot depicts different values of
the electric charge, where we have used the following values $\{M=10.0,\,q_m=0.75\}$. In the right plot, we have varied the values of the magnetic charge, and considered the values of the constants as follows $\{M=10.0,\,q_e=0.5\}$.  } 
\label{fig_Fqe_SV}
\end{figure*}

In the first scenario, we present three curves for different values of the electric charge ($q_e=0.25$, $q_e=0.5$ and $q_e=0.75$), as illustrated in the left plot of Fig. \ref{fig_Fqe_SV}. In the second scenario, we highlight  the behavior of this scalar when varying the magnetic charge for three different curves with ($q_m=0.25$, $q_m=0.5$ and $q_m=0.75$), as shown in the right plot of Fig. \ref{fig_Fqe_SV}.
We observe in the left plot of Fig.~\ref{fig_Fqe_SV} that as the electric charge varies, the amplitude decreases, while in the right plot of Fig.~\ref{fig_Fqe_SV}, the amplitude is greater when $q_e = q_m$.

By inverting the expression \eqref{field_camp}, we obtain $r(\cal{\varphi})$ and determine
\begin{align}
& V(\varphi) = \frac{4M\left(q_{e}^{2}+q_{m}^{2}\right)}{5\kappa^{2}\Big[\left(q_{e}^{2}+q_{m}^{2}\right)\text{sech}^{2}\left(\kappa\text{\ensuremath{\varphi}}\sqrt{\epsilon}\right)\Big]^{5/2}}.\label{V2_SV}
\end{align}  
If we take $q_e=0$ in this potential, we recover the potential obtained in \cite{Rodrigues2023}.

\begin{figure*}[th!]
\includegraphics[scale=0.57]{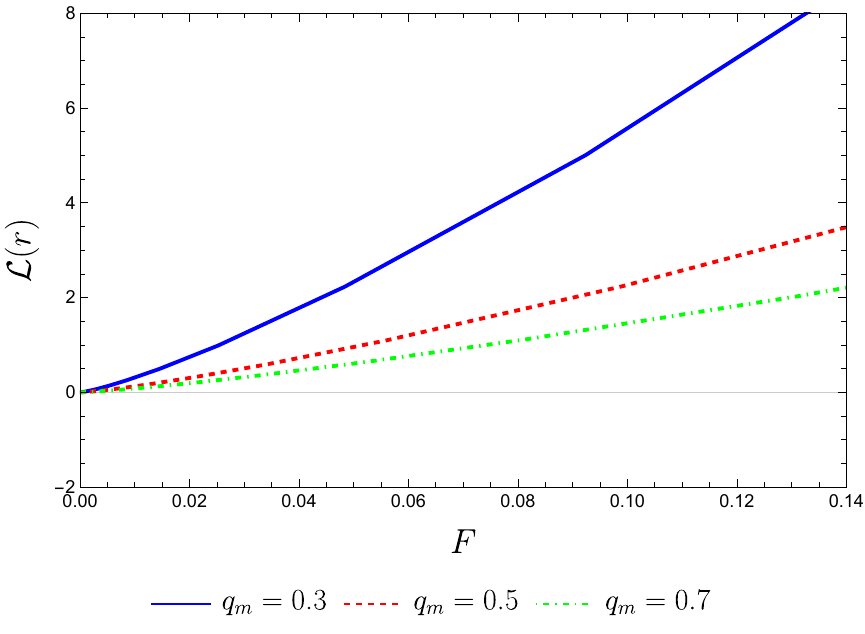}
	\hspace{0.75cm}
\includegraphics[scale=0.57]{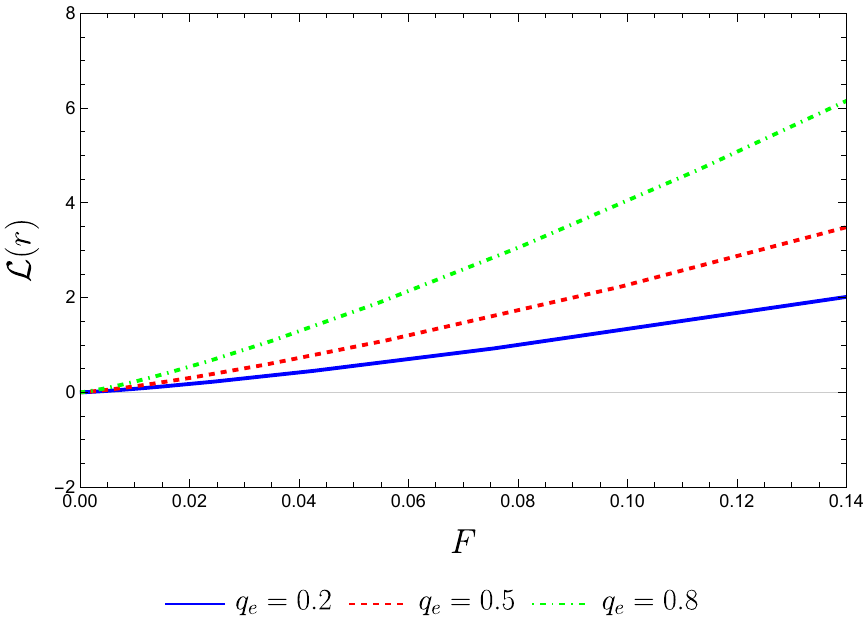}
\caption{The NLED Lagrangian ${\cal L}(F) $, for different values of the electric charge and the magnetic charge. (i) In the left plot, we use the values of the constants as follows $\{M=10.0,\,q_e=0.5\}$. (ii) In the rigth plot, we consider $\{M=10.0,\,q_m=0.5\}$.} 
\label{LxF_SV}
\end{figure*}

\begin{figure*}[ht!]
\includegraphics[scale=0.57]{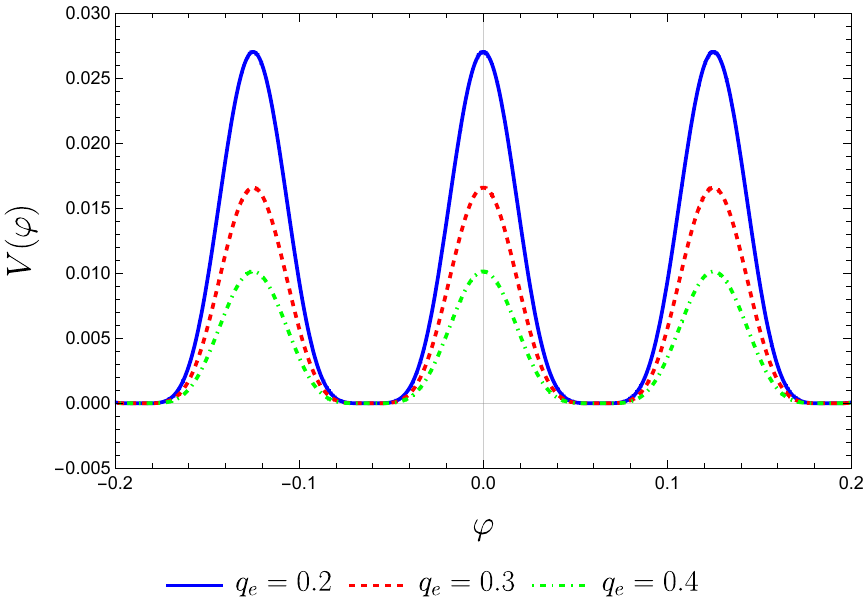}
	\hspace{0.75cm}
\includegraphics[scale=0.57]{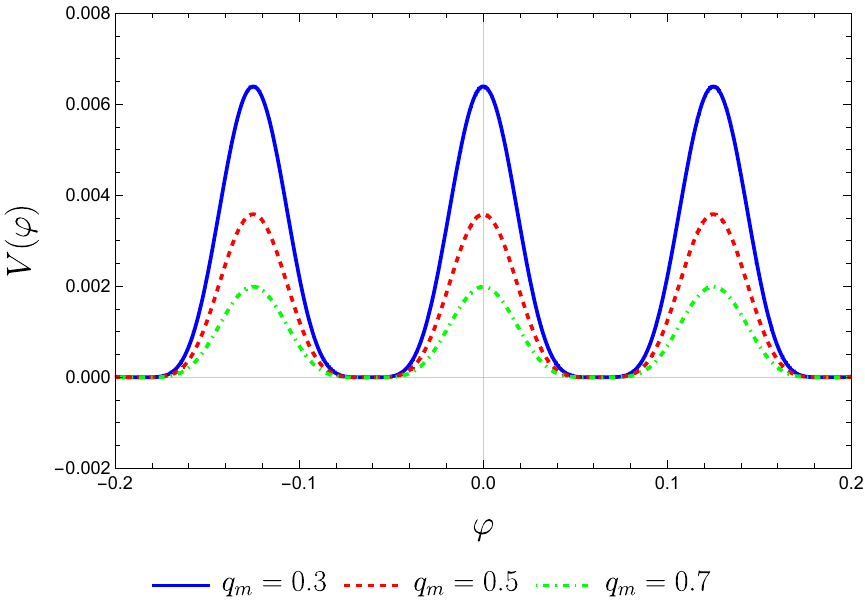}
\caption{The scalar field potential $V(\varphi)$, described by Eq.~\eqref{V2_SV} for different values of the electric charge and the magnetic charge, taking $\epsilon=-1$. (i) The left plot depicts the values of the constants $\{M=1.0,\,q_m=0.3\}$. (ii) The right plot represents the values for $\{M=1.0,\,q_e=0.5\}$.} 
\label{Vxphi_SV}
\end{figure*}

\begin{figure}[ht!]
\includegraphics[scale=0.57]{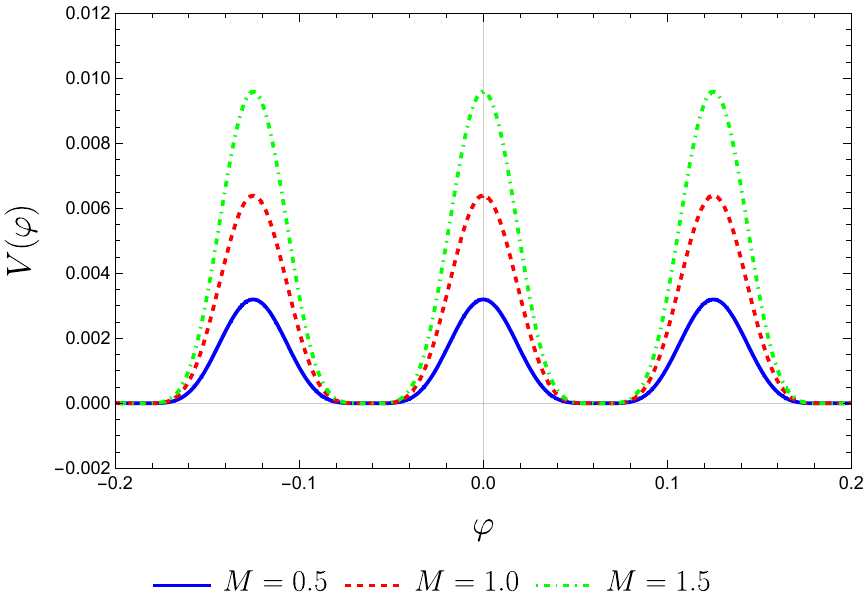}
\caption{The scalar field potential $V(\varphi)$, described by Eq.~\eqref{V2_SV} for different values of the mass, taking $\epsilon=-1$,  for the values of the constants $\{q_e=0.3,\,q_m=0.5\}$ .} 
\label{VxphiMass_SV}
\end{figure}

In the plots of Fig. \ref{LxF_SV} we show the behavior of the Lagrangian density of NLED as a function of $F$ using a parametric plot, for different values of the electric charge and of the magnetic charge. We also illustrate the behavior of the scalar field potential given by Eq. \eqref{V2_SV} in three distinct plots, each showing three curves with variations of electric charge, magnetic charge and mass values, as detailed in Figs. \ref{Vxphi_SV} and \ref{VxphiMass_SV}. We notice that the amplitude of the potential also varies as we change the values for the electric charge, magnetic charge and mass.

\section{Third Dyon solution.}\label{solIII}

\subsection{Metric function and horizons}

In Eq.~\eqref{L_BB}, to ensure that the Lagrangian ${\cal L}_{\text{NLED}}(r)$ remains real, the expression inside the square root in Eq.~\eqref{L_BB} must always be a positive real number. To satisfy this condition, we impose the following requirement:
\begin{eqnarray} &&\left[\Sigma(r)^{2}A''(r)-2A(r)\left(\Sigma(r)\Sigma''(r)+\Sigma'(r)^{2}\right)+2\right]^{2} =
\nonumber \\
&& \qquad \qquad = 64q_{e}^{2}\,q_{m}^{2}.
\end{eqnarray}
By considering the function \eqref{Sig1}, we simplify the previous expression as
\begin{align}
&\left[A''(r)\left(q_{e}^{2}+q_{m}^{2}+r^{2}\right)-2A(r)+2\right]^{2}=64q_{e}^{2}q_{m}^{2}\,, \label{eqAIII}
\end{align}
which yields the following metric function
\begin{eqnarray}
    A(r)&=&1-4\sqrt{q_{e}^{2}\,q_{m}^{2}}+c_{1}\left(q_{e}^{2}+q_{m}^{2}+r^{2}\right)+\frac{c_{2}\,r}{2\left(q_{e}^{2}+q_{m}^{2}\right)}
    \nonumber\\
&&+\frac{c_{2}\left(q_{e}^{2}+q_{m}^{2}+r^{2}\right)}{2\left(q_{e}^{2}+q_{m}^{2}\right)^{3/2}}\, \tan^{-1}\left(\frac{r}{\sqrt{q_{e}^{2}+q_{m}^{2}}}\right).\label{aIII}
\end{eqnarray}
The integration constants $c_1$ and $c_2$ are related to the gravitational mass or the gravitational energy of the system. We will discuss this relationship  in more detail when we calculate the Komar mass for this model.

\begin{figure*}[ht!]
\includegraphics[scale=0.56]{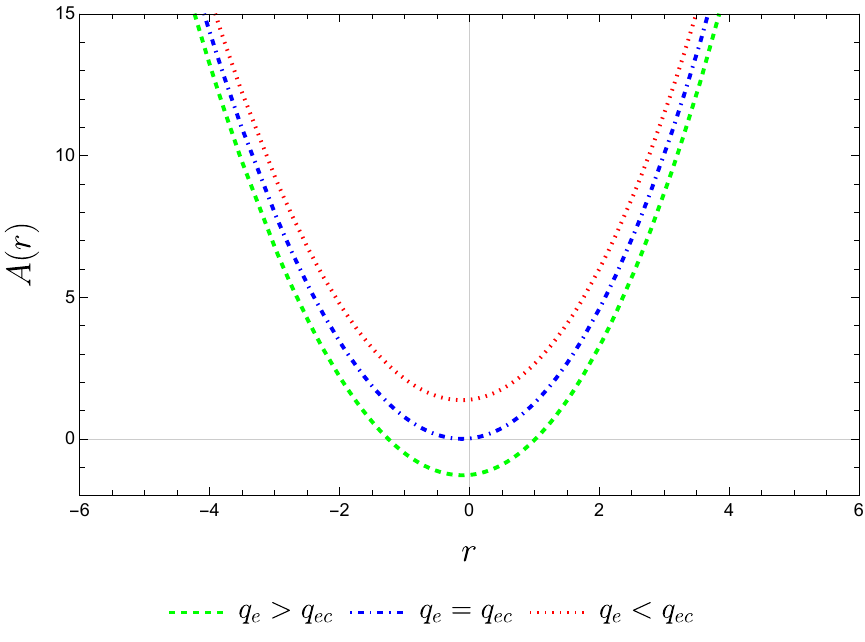}
	\hspace{0.75cm}
\includegraphics[scale=0.56]{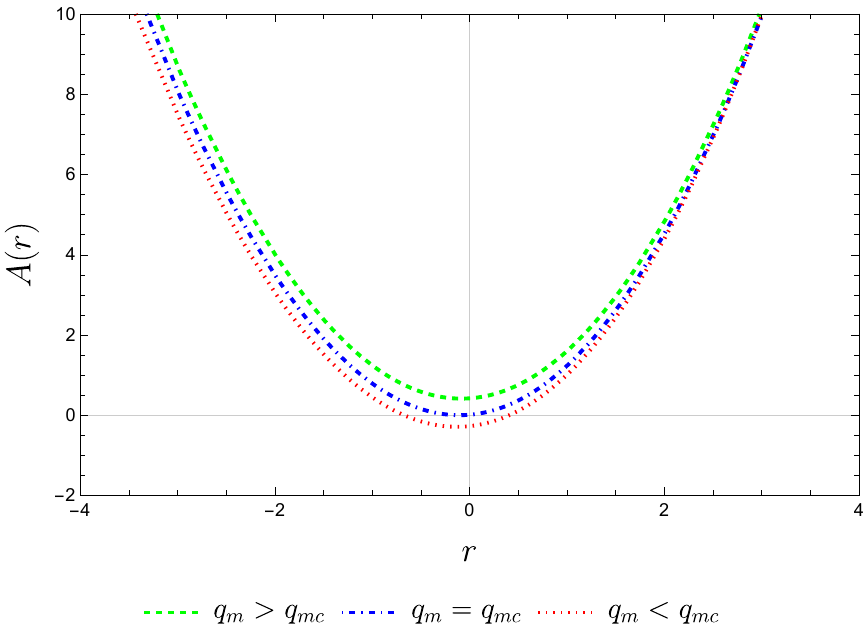}
\caption{The  metric function $A(r) $, given by Eq.~\eqref{aIII} for electric charge and magnetic charge scenarios. (i) In the left plot, we have used the values of the constants as follows $\{q_m=2.0,\,c_1=1.0, \,c_2=1.0\}$. (ii) In the right plot, we have considered the following vlaues $\{q_m=0.7,\,c_1=1.0, \,c_2=1.0\}$.} 
\label{aqe_III}
\end{figure*}

In this case, we solve Eqs. \eqref{rH} and \eqref{der_a} simultaneously to determine the critical electric charge, which for this solution is $q_{ec}=0.681$ when the constants are assigned the values $q_m=2.0$, $c_1=1$ and $c_2=1$. In the left plot of Fig. \ref{aqe_III} we show the behavior of the metric function \eqref{aIII} with respect to the radial coordinate $r$ for three different scenarios of the electric charge: $q_e > q_{ec}$, $q_e = q_{ec}$ and $q_e < q_{ec}$.
If the electric charge is greater than the critical value $q_e > q_{ec}$, no horizon is formed, which leads to a wormhole geometry with a traversable throat at $r=0$.
When the electric charge is equal to the critical value $q_e = q_{ec}$, we observe a degenerate double horizon at $r=0$, with a bounce still present at $r=0$.
Finally, when the electric charge is smaller than the critical electric charge, $q_e < q_{ec}$, we observe the presence of two event horizons. At $r = 0$, a bounce occurs, allowing a transition between the regions $r > 0$ and $r < 0$, and vice versa, due to the symmetry with respect to $r$.  
Furthermore, we observe that varying the values of the solution parameters enhances the asymmetrical nature of the metric function \eqref{aIII}.

We used the same procedure developed earlier to calculate $q_{ec}$, but now to determine the critical magnetic charge. We found that $q_{mc}=2.093$ when the values of the constants are $q_e=0.7$, $c_1=1$, and $c_2=1$. In the right plot of Fig. \ref{aqe_III}, we show the behavior of the metric function \eqref{aIII} with respect to the radial coordinate $r$ for three different scenarios of the magnetic charge: $q_m > q_{mc}$, $q_m = q_{mc}$, and $q_m < q_{mc}$.

In general, the solution given by Eq.\;\eqref{aIII} is not asymptotically flat, which complicates the definition of the mass $M$ or the gravitational energy. However, this limitation can be overcome by making an appropriate choice of constants to render the metric function \eqref{aIII} asymptotically flat, thus allowing for an accurate determination of the mass or gravitational energy of the system. To achieve this goal, we define the constant $c_2$ as
\begin{equation}
c_2 = -\frac{4c_{1}\left(q_{e}^{2} + q_{m}^{2}\right)^{3/2}}{\pi}.\label{c2_III}
\end{equation}

With this choice, the metric function of the model becomes asymptotically flat. After substituting the constant $c_2$, the metric function \eqref{aIII} can now be rewritten as
\begin{eqnarray}
&A(r)=1+c_{1}\left(r^2 + q_{e}^{2}+q_{m}^{2}   -\frac{2r\sqrt{q_{e}^{2}+q_{m}^{2}}}{\pi}\right)
 -4\sqrt{q_{e}^{2}+q_{m}^{2}}
   \nonumber\\
&-\frac{2c_{1}}{\pi}\left(q_{e}^{2}+q_{m}^{2}+r^{2}\right)\tan^{-1}\left(\frac{r}{\sqrt{q_{e}^{2}+q_{m}^{2}}}\right).\label{aIII2}
\end{eqnarray}

We also determine the Komar mass, as given by Eq.~\eqref{m_Komar2}, for the metric function described by Eq.~\eqref{aIII2}, which for this model is given by:  
\begin{align}
M_K = \frac{2\bar{{c}}_{1}}{3\pi}, \label{Kom_A3}
\end{align}
where we replace $c_{1}=-\bar{c}_{1}\left(q_{e}^{2}+q_{m}^{2}\right)^{-3/2}$, and $\bar{c}_1$ is a new positive real numerical constant that is independent of the electric and magnetic charges, ensuring that the Komar mass is also independent. 

Instead of assuming a constant $c_2$, it is also possible to adopt the constant $c_1$. In this way, we now consider
\begin{equation}
c_1=-\frac{\pi c_{2}}{4\left(q_{e}^{2}+q_{m}^{2}\right)^{3/2}},
\end{equation} 
so that the metric function \eqref{aIII} becomes asymptotically flat. Substituting this constant $c_1$ into the metric function \eqref{aIII}, we obtain:
\begin{eqnarray}
A(r)&=&1-\frac{\pi c_{2}\left(q_{e}^{2}+q_{m}^{2}+r^{2}\right)}{4\left(q_{e}^{2}+q_{m}^{2}\right)^{3/2}}+\frac{c_{2}r}{2\left(q_{e}^{2}+q_{m}^{2}\right)}
-4\sqrt{q_{e}^{2}\,q_{m}^{2}}
\nonumber\\
&&	+\frac{c_{2}\left(q_{e}^{2}+q_{m}^{2}+r^{2}\right)}{2\left(q_{e}^{2}+q_{m}^{2}\right)^{3/2}}	\, \tan^{-1}\left(\frac{r}{\sqrt{q_{e}^{2}+q_{m}^{2}}}\right)
    .\label{aIV3}
\end{eqnarray}

In this manner, we determine the Komar mass, as given by Eq. \eqref{m_Komar2}, for the metric function \eqref{aIV3}. Its explicit form is now
\begin{align}
M_K=\frac{c_2}{6}.\label{Kom2_A4}
\end{align}

\subsection{Kretschmann and eletromagnetic scalar}

\begin{figure}[th!]
\includegraphics[scale=0.55]{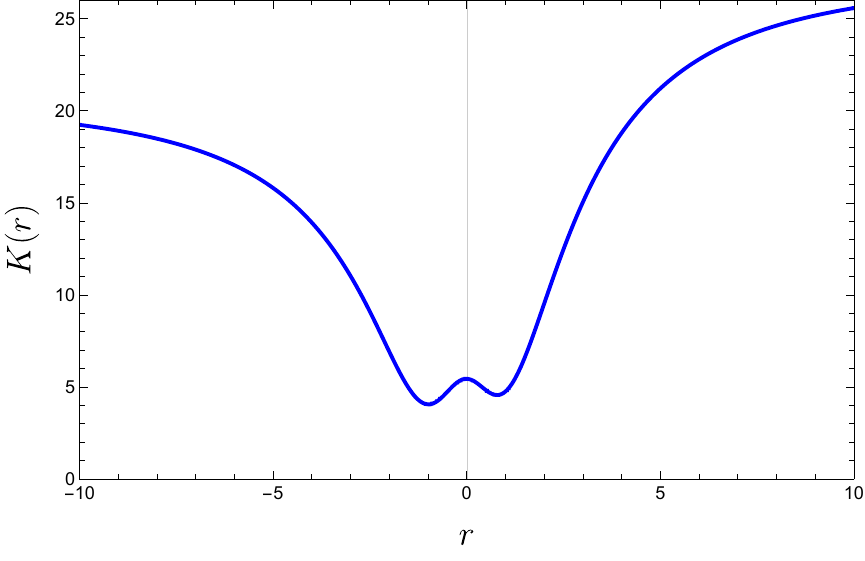}
\caption{The behavior of the Kretschmann scalar $K(r) $. We have used the values of the constants as follows $\{q_e=1.0,\,q_m=2.0,\,c_1=1.0, \,c_2=1.0\}$. } 
\label{fig_K}
\end{figure}

In order to gain a better understanding of the regularity properties of our model, we calculate the Kretschmann scalar for this solution.
Taking the limit of $r\rightarrow 0$ for the Kretschmann scalar, we observe that it is regular under this condition
\begin{eqnarray}
&&\lim_{r\rightarrow0}K\left(r\right)=4\Big[3c_{1}^{2}\left(q_{e}^{2}+q_{m}^{2}\right)^{2}+32q_{e}^{2}\,q_{m}^{2}-16\sqrt{q_{e}^{2}q_{m}^{2}}
\nonumber\\
&&
-4c_{1}\left(q_{e}^{2}+q_{m}^{2}\right)\left(4\sqrt{q_{e}^{2}\,q_{m}^{2}}-1\right)+3\Big]\big/ \left(q_{e}^{2}+q_{m}^{2}\right)^{2}.
\end{eqnarray}
We also observe that the Kretschmann scalar is regular for very large values of $r$
\begin{equation}
\lim_{r\rightarrow\infty}K\left(r\right)=\frac{3\left(4c_{1}\left(q_{e}^{2}+q_{m}^{2}\right)+\pi c_{2}\sqrt{\frac{1}{q_{e}^{2}+q_{m}^{2}}}\right)^{2}}{2\left(q_{e}^{2}+q_{m}^{2}\right)^{2}}.
\end{equation}
The Kretschmann scalar expression for this model is quite complicated, therefore, we illustrate its regular behavior as depicted in Fig. \ref{fig_K}.

\subsection{Matter Lagrangian and Potential }

Substituting Eq.\,\eqref{Sig1} and the metric function \eqref{aIII} into Eq.~\eqref{V}, we obtain the form of the potential as a function of the radial coordinate:
\begin{eqnarray}
    V(r) &=&-\frac{3c_{2}r-4c_{1}\left(q_{e}^{2}+q_{m}^{2}\right)^{2}}{2\kappa^{2}\left(q_{e}^{2}+q_{m}^{2}\right)\left(q_{e}^{2}+q_{m}^{2}+r^{2}\right)}
\nonumber\\
&&-\frac{c_{2}\left(q_{e}^{2}+q_{m}^{2}+3r^{2}\right)\tan^{-1}\left(\frac{r}{\sqrt{q_{e}^{2}+q_{m}^{2}}}\right)}{2\kappa^{2}\left(q_{e}^{2}+q_{m}^{2}\right)^{3/2}\left(q_{e}^{2}+q_{m}^{2}+r^{2}\right)}	
	.\label{V_III}
\end{eqnarray}

Having used the appropriate quantities such as the function \eqref{Sig1}, the metric function \eqref{aIII}, the scalar field \eqref{field_camp} and the potential \eqref{V_III}, we can explicitly express the Lagrangian ${\cal L}_{\textrm{NLED}}$ for this model, which is derived from the general Lagrangian given by Eq. \eqref{L_BB},
\begin{eqnarray}
&&{\cal L}_{\textrm{NLED}}(r) =-\frac{3c_{1}}{2}+\frac{\sqrt{q_{e}^{2}\,q_{m}^{2}}}{\left(q_{e}^{2}+q_{m}^{2}+r^{2}\right)^{2}}\bigg[\left(q_{e}^{2}+q_{m}^{2}+r^{2}\right)	
\nonumber\\
&& \qquad
+\sqrt{\left(q_{e}^{2}+q_{m}^{2}+r^{2}-1\right)\left(q_{e}^{2}+q_{m}^{2}+r^{2}+1\right)}\bigg]	,
\label{L2_BB} 
\end{eqnarray}
whose derivative ${\cal L}_F$, Eq. \eqref{LF_BB}, is given by
\begin{eqnarray}
&& {\cal L}_F(r) =\,\frac{\sqrt{q_{e}^{2}\,q_{m}^{2}}}{q_{m}^{2}}\bigg[\left(q_{e}^{2}+q_{m}^{2}+r^{2}\right)
\nonumber\\
&& \qquad
+\sqrt{\left(q_{e}^{2}+q_{m}^{2}+r^{2}-1\right)\left(q_{e}^{2}+q_{m}^{2}+r^{2}+1\right)} \bigg].\label{LF2_BB}
\end{eqnarray}

Now, with the function \eqref{Sig1}, the metric function \eqref{aIII}, the scalar field \eqref{field_camp}, the potential \eqref{V_III}, the Lagrangian \eqref{L2_BB}, and the derivative of the Lagrangian \eqref{LF2_BB}, we can satisfy all equations \eqref{sol2}–\eqref{sol3} and \eqref{RC}, as well as all components of the equations of motion \eqref{EqF00}–\eqref{EqF22}.

We note that for small values of $r$ the Lagrangian \eqref{L2_BB} and the derivative \eqref{LF2_BB} are pure imaginary if $q_{e}^{2}+q_{m}^{2}\ll1$. 

In turn, from Eq.~\eqref{F2}, we determined that the electromagnetic scalar for this solution has the following form:
\begin{eqnarray}
&&F(r)=
\frac{1}{q_{e}^{2}+q_{m}^{2}+r^{2}}-\frac{q_{e}^{2}+r^{2}}{\left(q_{e}^{2}+q_{m}^{2}+r^{2}\right)^{2}}-q_{m}^{2}
\nonumber\\
&& \quad
+\frac{q_{m}^{2}\sqrt{\left(q_{e}^{2}+q_{m}^{2}+r^{2}-1\right)\left(q_{e}^{2}+q_{m}^{2}+r^{2}+1\right)}}{\left(q_{e}^{2}+q_{m}^{2}+r^{2}\right)}.\label{FIII}
\end{eqnarray}

\begin{figure*}[ht!]
	\centering
	\includegraphics[scale=0.56]{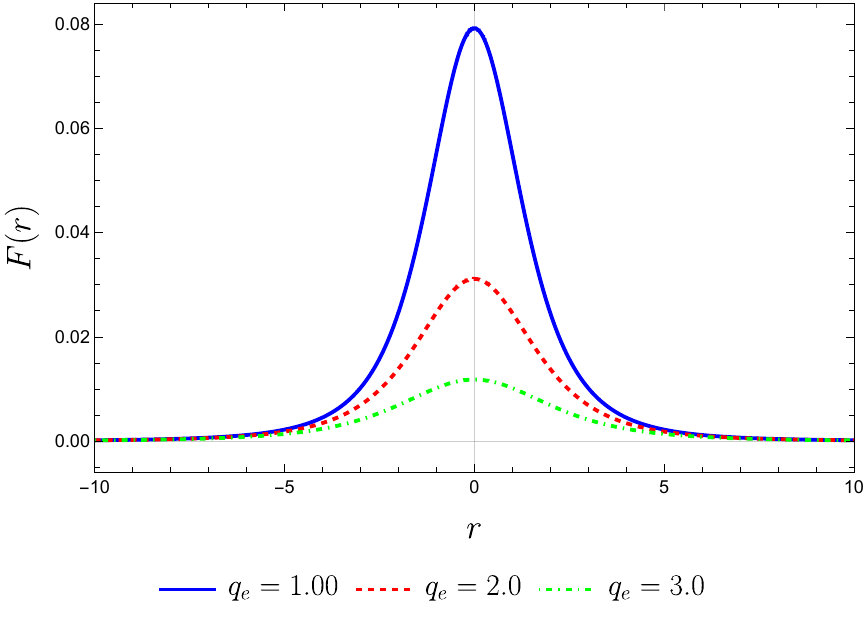}
	\hspace{0.75cm}
	\includegraphics[scale=0.56]{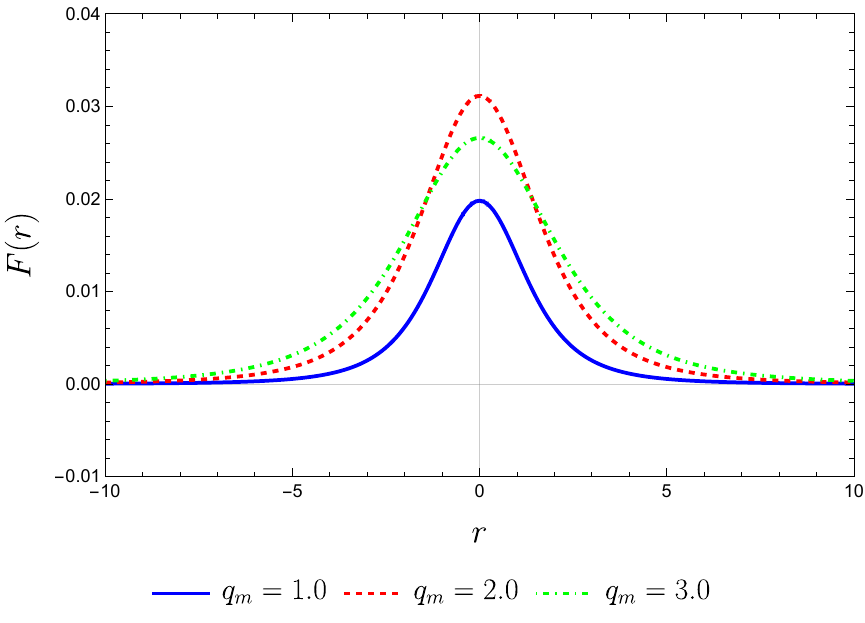}
	\caption{The electromagnetic scalar $F$, given by Eq.~\eqref{FIII}, for different values of
		the electric and magnetic charges.  (i) In the left plot, we use the values of the constants as follows $\{q_m=2.0,\,c_1=1.0, \,c_2=1.0\}$. (ii) In the right plot, we consider $\{q_e=2.0,\,c_1=1.0, \,c_2=1.0\}$.} 
	\label{fig_Fqe_III}
\end{figure*}

Once more, we describe the behavior of the electromagnetic scalar \eqref{FIII} in two different scenarios. First, we present three curves for different values of electric charge ($q_e=1.0$, $q_e=2.0$ and $q_e=3.0$), as shown in the left plot of Fig. \ref{fig_Fqe_III}. Secondly, we illustrate the behavior of this scalar when varying the magnetic charge for three different curves with ($q_m=1.0$, $q_m=2.0$ and $q_m=3.0$), as shown in the right plot of Fig.  \ref{fig_Fqe_III}.

By inverting Eqs. \eqref{field_camp} and \eqref{FIII}, we rewrite $r(\cal{\varphi})$ and $r(F)$ respectively. In this way, we express the Lagrangian in terms of $F$ in the form 
\begin{eqnarray}
   {\cal L} _{\rm NLED} (F) =	&-\frac{3c_{1}}{2}+\frac{\left(2Fq_{m}^{2}-\sqrt{q_{m}^{8}-4Fq_{m}^{6}}+q_{m}^{4}\right)}{2q_{m}^{4}}\times
   \nonumber\\
   &
   \times \sqrt{q_{e}^{2}\,q_{m}^{2}}\left[\sqrt{\frac{2q_{m}^{4}}{2Fq_{m}^{2}-\sqrt{q_{m}^{8}-4Fq_{m}^{6}}+q_{m}^{4}}}\right.
   \nonumber\\
   &\left.
+\sqrt{\frac{\left(-2Fq_{m}^{2}+\sqrt{q_{m}^{8}-4Fq_{m}^{6}}+q_{m}^{4}\right)}{2Fq_{m}^{2}-\sqrt{q_{m}^{8}-4Fq_{m}^{6}}+q_{m}^{4}}}\,\,\right],
 \label{L3_BB} 
\end{eqnarray}   
and the potential in terms of the scalar field as
\begin{eqnarray}
& V(\varphi) = \frac{c_{2}\tan^{-1}\left[\tan\left(\varphi\sqrt{\kappa^{2}(-\epsilon)}\right)\right]\left[\cosh\left(2\kappa\varphi\sqrt{\epsilon}\right)-2\right]}{2\kappa^{2}\left(q_{e}^{2}+q_{m}^{2}\right)^{3/2}}
\nonumber\\
& + \frac{\cosh^{2}\left(\kappa\varphi\sqrt{\epsilon}\right)\left[4c_{1}\left(q_{e}^{2}+q_{m}^{2}\right)^{3/2}-3c_{2}\tan\left(\varphi\sqrt{\kappa^{2}(-\epsilon)}\right)\right]}{2\kappa^{2}\left(q_{e}^{2}+q_{m}^{2}\right)^{3/2}}.\label{V3_BB}
\end{eqnarray}  

\begin{figure}[ht!]
\includegraphics[scale=0.56]{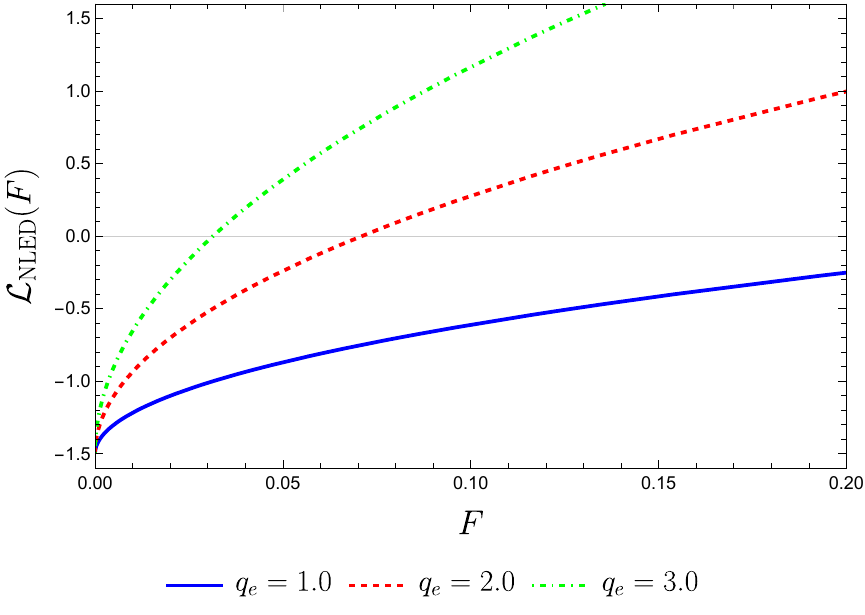}
\caption{The NLED Lagrangian ${\cal L}(F) $, given by Eq.~\eqref{L3_BB} for different values of the electric charge.  We have used the values of the constants as follows $\{q_m=2.0,\,c_1=1.0, \,c_2=1.0\}$.  } 
\label{LxFIII_BB}
\end{figure}

\begin{figure*}[ht!]
\includegraphics[scale=0.57]{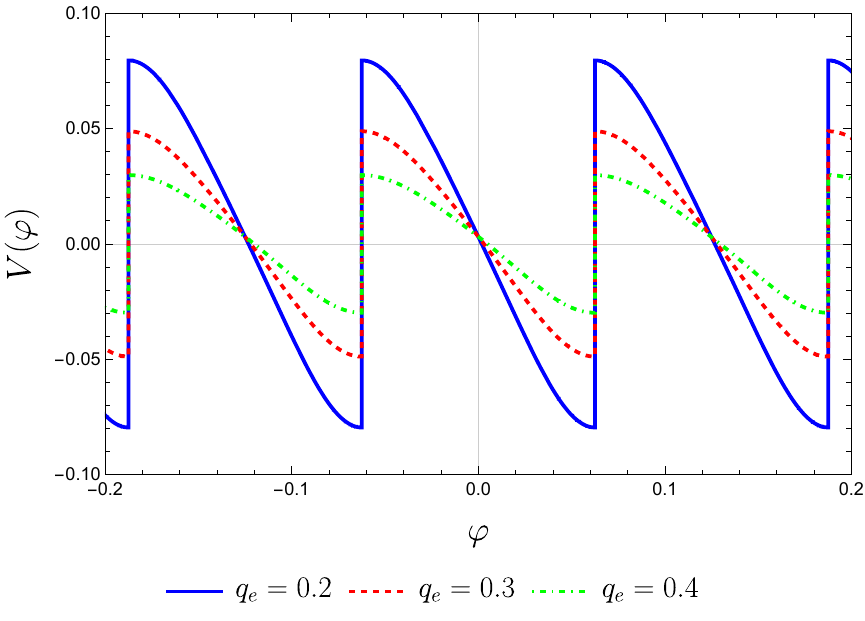}
	\hspace{0.75cm}
\includegraphics[scale=0.57]{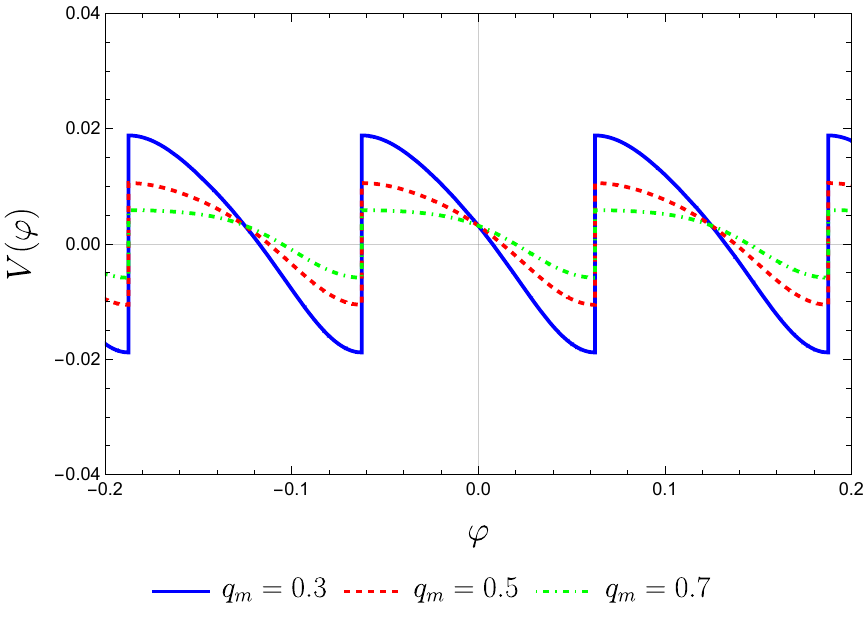}
\caption{The scalar field potential $V(\varphi)$, described by Eq.~\eqref{V3_BB} for different values of the electric and magnetic charges, taking $\epsilon=-1$. (i) I the left plot, we consider values $\{q_m=0.3,\,c_1=1.0, \,c_2=1.0\}$. In the right plot, we use the values $\{q_e=0.5,\,c_1=1.0, \,c_2=1.0\}$.} 
\label{VxphiIII_BB}
\end{figure*}

In Fig.\;\ref{LxFIII_BB} we show the behavior of the Lagrangian density ${\cal L} _{\rm NLED}$ as a function of $F$, for different values of the electric charge. We also illustrate the behavior of the scalar field potential given by Eq.\,\eqref{V3_BB} in distinct plots, each showing three curves with variations of electric charge and magnetic charge, as detailed in Figs. \ref{VxphiIII_BB}. We notice that the amplitude of the potential also varies as we change the values for electric charge, magnetic charge and mass.

Additionally, we observe that for very small values of $F$ we get 
\begin{equation}
{\cal L} _{\rm NLED} (F) \sim -\frac{3c_{1}}{2}+\frac{2\sqrt{F}}{q_{m}^{2}}\left(\frac{\sqrt{q_{e}^{2}q_{m}^{4}}}{\sqrt{2}}+\frac{\sqrt{q_{m}^{2}}\sqrt{q_{2}^{2}q_{m}^{2}}}{\sqrt{2}}\right)
.
\end{equation}

\section{Fourth Dyon solution}\label{solIV}
\subsection{Metric function and horizons}

In this section, we consider the function described by \eqref{Sig1} and a generalization of the metric function \eqref{aIII} described by:
\begin{align}
    A(r)=&1-\frac{2M}{\sqrt{q_{e}^{2}+q_{m}^{2}+r^{2}}}-4\sqrt{q_{e}^{2}\,q_{m}^{2}}
    \nonumber\\
&+c_{1}q_{e}^{2}+c_{1}\left(q_{m}^{2}+r^{2}\right)+\frac{c_{2}r}{2\left(q_{e}^{2}+q_{m}^{2}\right)}
\nonumber\\
&
+\frac{c_{2}\left(q_{e}^{2}+q_{m}^{2}+r^{2}\right)\tan^{-1}\left(\frac{r}{\sqrt{q_{e}^{2}+q_{m}^{2}}}\right)}{2\left(q_{e}^{2}+q_{m}^{2}\right)^{3/2}}.\label{aIV}
\end{align}
where $M$ denotes the mass of the black hole.

\begin{figure*}[th!]
\includegraphics[scale=0.56]{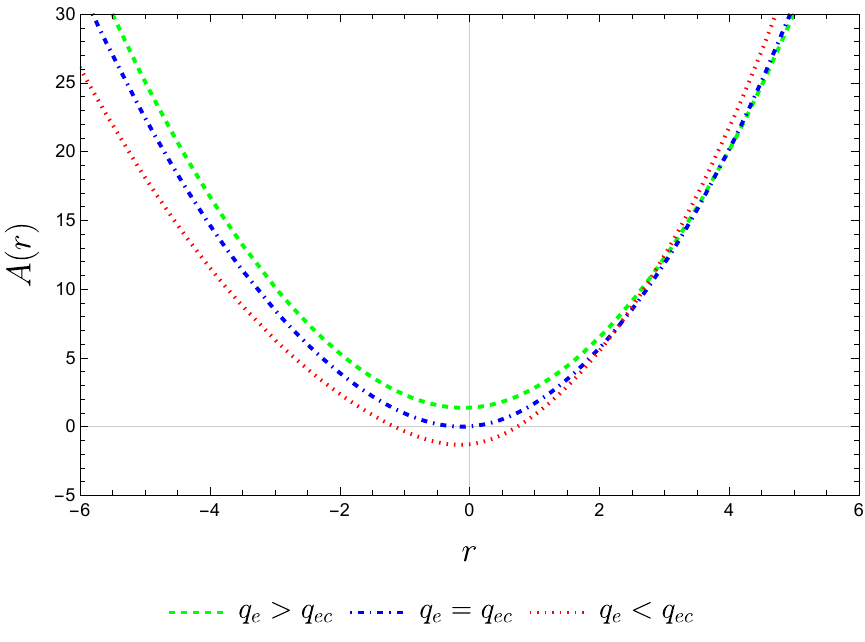}
	\hspace{0.75cm}
\includegraphics[scale=0.56]{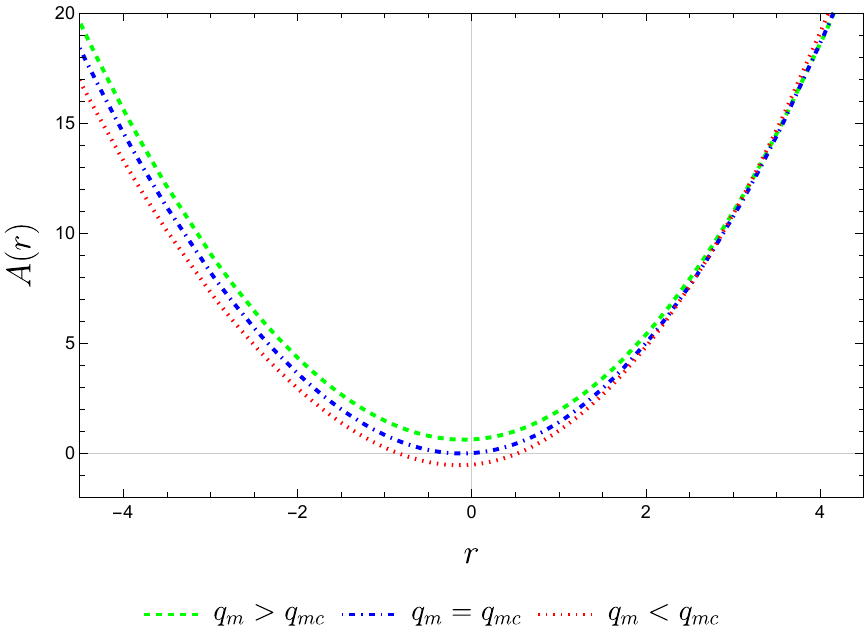}
\caption{The  metric function $A(r) $, given by Eq.~\eqref{aIV} for three electric charge scenarios. We have used the values of the constants as follows $\{M=2.0,\,q_m=0.25,\,c_1=1.0, \,c_2=1.0\}$. } 
\label{figa2}
\end{figure*}

We solve Eqs. \eqref{rH} and \eqref{der_a} simultaneously to determine the critical electric charge using the metric function \eqref{aIV}, which has the value $q_e=1.725$ if we set the constants $M=2.0$, $q_m=0.25$, $c_1=1$, and $c_2=1$. In the left plot of Fig. \ref{figa2} we show the behavior of the metric function \eqref{aIV} with respect to the radial coordinate $r$ for three different scenarios of the electric charge: $q_e > q_{ec}$, $q_e = q_{ec}$ and $q_e < q_{ec}$. We note that the behavior in these cases is similar to Fig. \ref{aqc_SV}. Moreover, we emphasize that the asymmetric nature of the metric function becomes clearer if we change the values of the solution parameters.

In the case of the critical magnetic charge, we obtain the value $q_{mc}=1.888$ by defining the constants $M=1.0$, $q_e=0.5$, $c_1=1$ and $c_2=1$. In the right plot of Fig. \ref{figa2} we show the behavior of the metric function \eqref{aIV} as a function of the radial coordinate $r$ for three different scenarios of the magnetic charge: $q_m > q_{mc}$, $q_m = q_{mc}$ and $q_m < q_{mc}$. We also note that the behavior in these cases is similar to Fig. \ref{aqc_SV}. In particular, the asymmetric nature of the metric function becomes clearer when we alter the values of the solution's parameters.

\begin{figure}[th!]
\includegraphics[scale=0.56]{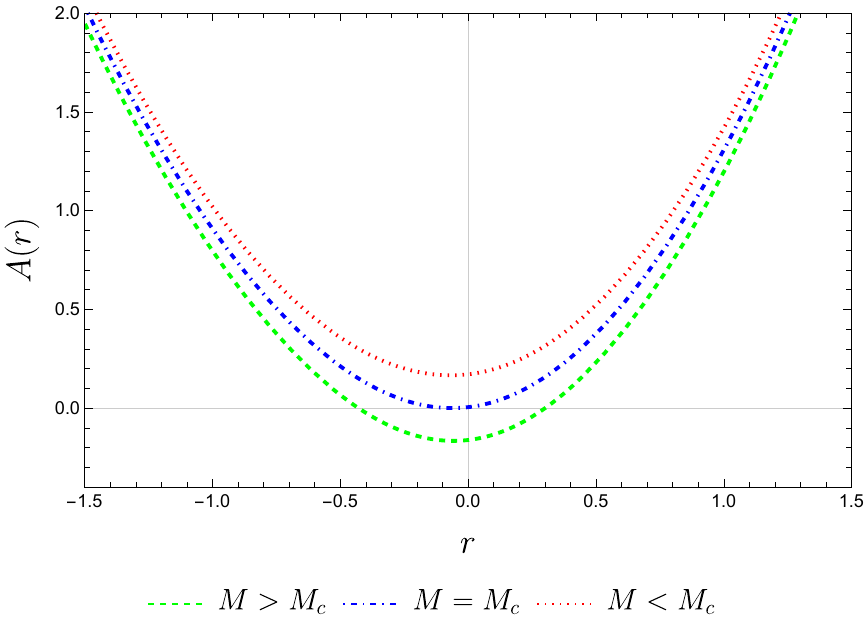}
\caption{The  metric function $A(r) $, given by Eq.~\eqref{aIV} for three mass scenarios. We have used the values of the constants as follows $\{q_e=0.5,\,q_m=0.75,\,c_1=1.0, \,c_2=0.12\}$. } 
\label{figa2c}
\end{figure}

Finally, the critical mass, which is determined by simultaneously solving Eqs (\ref{rH}) and (\ref{der_a}), takes the value $M_c = 0.138$ if we choose the constant values $q_e=0.5$, $q_m=0.75$, $c_1=1$ and $c_2=0.12$. Figure \ref{figa2c} shows the behavior of the metric function \eqref{aIV} as a function of the radial coordinate $r$ for three mass scenarios: $M > M_c$, $M = M_c$, and $M < M_c$. We note that the behavior in these cases is similar to that in the right plot of Fig. \ref{aqc_SV}.

Once again, we note that the metric function \eqref{aIV} is not asymptotically flat. However, by defining
\begin{equation}
c_2=-\frac{4c_{1}\left(q_{e}^{2}+q_{m}^{2}\right)^{3/2}}{\pi},
\end{equation} 
the Eq. \eqref{aIV} becomes asymptotically flat. Substituting this constant $c_2$ into \eqref{aIV}, we obtain
\begin{eqnarray}
A(r)&=&1-\frac{2M}{\sqrt{q_{e}^{2}+q_{m}^{2}+r^{2}}}-4\sqrt{q_{e}^{2}+q_{m}^{2}}
    \nonumber\\
&& +c_{1}\left(-\frac{2r}{\pi} \sqrt{q_{e}^{2}+q_{m}^{2}} +q_{e}^{2}+q_{m}^{2}+r^{2}\right)
\nonumber\\
&& 
-\frac{2c_{1}}{\pi}\left(q_{e}^{2}+q_{m}^{2}+r^{2}\right)\tan^{-1}\left(\frac{r}{\sqrt{q_{e}^{2}+q_{m}^{2}}}\right).
	\label{aIV2}
\end{eqnarray}

Computing the Komar mass via Eq. \eqref{m_Komar2}, for the metric function \eqref{aIV2}, gives
\begin{align}
M_K=M+\frac{2\bar{c}_{1}}{3\pi},\label{Kom_A4}
\end{align}
where we again substitute $c_{1}=-\frac{\bar{c}_{1}}{\left(q_{e}^{2}+q_{m}^{2}\right)^{3/2}}$, and $\bar{c}_1$ is a new positive real constant that is independent of the electric and magnetic charges, so that the Komar mass is also independent of these quantities.

\begin{figure}[th!]
\includegraphics[scale=0.56]{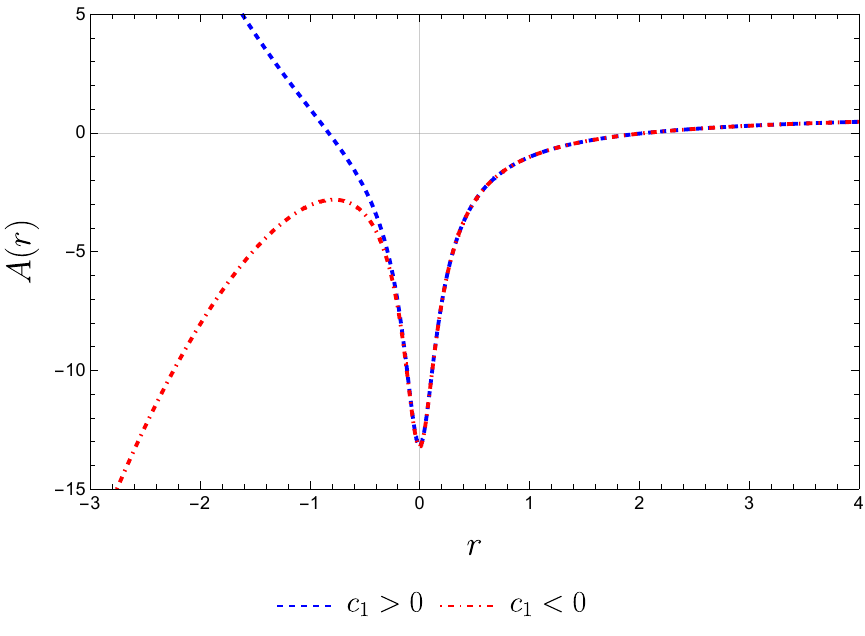}
\caption{The  metric function $A(r) $, given by Eq.~\eqref{aIV2}. We have used the values of the constants as follows $\{q_e=0.1,\,q_m=0.1$ and $M=1.0\}$. } 
\label{fig_aIV2}
\end{figure}

In Fig.\,\ref{fig_aIV2}, we illustrate the behavior of the metric function \eqref{aIV2} with two curves. We fix the constants $q_e=0.1$, $q_m=0.1$ and $c_2=1.0$ for both curves, and vary $c_1=\pm 1$ to plot the curves with positive and negative $c_1$, respectively.

\subsection{Kretschmann scalar}

\begin{figure}[ht!]
\includegraphics[scale=0.56]{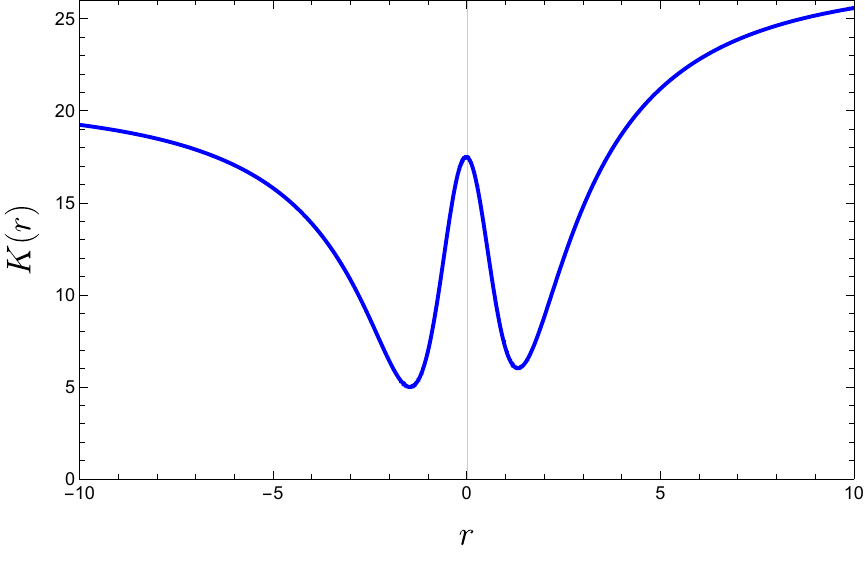}
\caption{The behavior of the Kretschmann scalar $K(r) $. We have used the values of the constants as follows $\{M=4.0,\,q_e=1.0,\,q_m=2.0,\,c_1=1.0, \,c_2=1.0\}$. }  \label{fig_K_IV}
\end{figure}

We have also calculated the Kretschmann scalar for this solution in order to better examine the regularity of our model. If we take the limit $r \rightarrow 0$ for the Kretschmann scalar, we find that it is regular in the $r\rightarrow 0$ limit
\begin{align}
&\lim_{r\rightarrow0}K\left(r\right)=\Bigg\{\left[2c_{1}\left(q_{e}^{2}+q_{m}^{2}\right)^{3}+2M\left(q_{e}^{2}+q_{m}^{2}\right)^{3/2}\right]^{2}
\nonumber
\\
&
+4\left(q_{e}^{2}+q_{m}^{2}\right)^{4}
+2\bigg[4M\left(q_{e}^{2}+q_{m}^{2}\right)^{3/2}-2\left(q_{e}^{2}+q_{m}^{2}\right)^{2}
\nonumber
\\
&
+\left(8\sqrt{q_{e}^{2}\,q_{m}^{2}}\left(q_{e}^{2}+q_{m}^{2}\right)^{3/2}-2c_{1}\left(q_{e}^{2}+q_{m}^{2}\right)^{5/2}\right)\times
\nonumber
\\
&
\qquad\times\sqrt{q_{e}^{2}+q_{m}^{2}}\bigg]^2
\Bigg\}\Big/\left(q_{e}^{2}+q_{m}^{2}\right)^{6},
\end{align}
while the $r\rightarrow\infty$ limit of the Kretschmann scalar for this solution is also regular
\begin{equation}
\lim_{r\rightarrow\infty}K\left(r\right)=\frac{3\Bigg[4c_{1}\left(q_{e}^{2}+q_{m}^{2}\right)+\pi c_{2}\sqrt{\frac{1}{q_{e}^{2}+q_{m}^{2}}}\Bigg]^{2}}{2\left(q_{e}^{2}+q_{m}^{2}\right)^{2}}.
\end{equation}
We illustrate the behavior of the Kretschmann scalar of this solution in Fig. \ref{fig_K_IV}.

\subsection{Matter Lagrangian and Potential reconstruction}

Substituting \eqref{Sig1} and the metric function \eqref{aIV} into the solution \eqref{V}, we determine the form of the potential as a function of the radial coordinate for this model, which is given by:
\begin{align}
&V(r)=-\frac{c_{2}\left(q_{e}^{2}+q_{m}^{2}+3r^{2}\right)\tan^{-1}\left(\frac{r}{\sqrt{q_{e}^{2}+q_{m}^{2}}}\right)}{2\kappa^{2}\left(q_{e}^{2}+q_{m}^{2}\right)^{3/2}\left(q_{e}^{2}+q_{m}^{2}+r^{2}\right)}	+
\nonumber\\
&\frac{\left(4c_{1}\left(q_{e}^{2}+q_{m}^{2}\right)^{2}-3c_{2}r\right)}{2\kappa^{2}\left(q_{e}^{2}+q_{m}^{2}\right)\left(q_{e}^{2}+q_{m}^{2}+r^{2}\right)}
+\frac{4m\left(q_{e}^{2}+q_{m}^{2}\right)}{5\kappa^{2}\left(q_{e}^{2}+q_{m}^{2}+r^{2}\right)^{5/2}}	
	.\label{V_IV}
\end{align}

Having used the appropriate quantities such as the function \eqref{Sig1}, the metric function \eqref{aIV}, the scalar field \eqref{field_camp} and the potential \eqref{V_IV}, we can explicitly express the Lagrangian ${\cal L}_{\textrm{NLED}}$ for this model, which is derived from the general Lagrangian, given by Eq.\,\eqref{L_BB},
\begin{align}
&{\cal L}_{\textrm{NLED}}(r) =-\frac{3c_{1}}{2}+\Bigg[\Bigg(\frac{9M^2\left(q_{e}^{2}+q_{m}^{2}\right)^{2}}{q_{e}^{2}+q_{m}^{2}+r^{2}}
\nonumber\\
&
+24M\sqrt{q_{e}^{2}\,q_{m}^{2}}\left(q_{e}^{2}+q_{m}^{2}\right)\sqrt{q_{e}^{2}+q_{m}^{2}+r^{2}}
\nonumber\\
&
+16q_{e}^{2}q_{m}^{2}\left(q_{e}^{2}+q_{m}^{2}+r^{2}-1\right)\left(q_{e}^{2}+q_{m}^{2}+r^{2}+1\right)\Bigg)^{\frac{1}{2}}
\nonumber\\
&
+4\sqrt{q_{e}^{2}\,q_{m}^{2}}\left(q_{e}^{2}+q_{m}^{2}+r^{2}\right)\Bigg]\Big/4\left(q_{e}^{2}+q_{m}^{2}+r^{2}\right)^{2}
\nonumber\\
&
-\frac{3M\left(q_{e}^{2}+q_{m}^{2}\right)}{20\left(q_{e}^{2}+q_{m}^{2}+r^{2}\right)^{5/2}},
\label{L2_BBb} 
\end{align}
and its derivative ${\cal L}_F$, given by Eq. \eqref{LF_BB}, takes the following form
\begin{align}
&{\cal L}_F(r) =\Bigg[\Bigg(24M\sqrt{q_{e}^{2}\,q_{m}^{2}}\left(q_{e}^{2}+q_{m}^{2}\right)\sqrt{q_{e}^{2}+q_{m}^{2}+r^{2}}
\nonumber\\
&
+16q_{e}^{2}q_{m}^{2}\left(q_{e}^{2}+q_{m}^{2}+r^{2}-1\right)\left(q_{e}^{2}+q_{m}^{2}+r^{2}+1\right)
\nonumber\\
&
+\frac{9M^{2}\left(q_{e}^{2}+q_{m}^{2}\right)^{2}}{q_{e}^{2}+q_{m}^{2}+r^{2}}\Bigg)^{\frac{1}{2}}+\frac{3M\left(q_{e}^{2}+q_{m}^{2}\right)}{\sqrt{q_{e}^{2}+q_{m}^{2}+r^{2}}}
\nonumber\\
&
+4\sqrt{q_{e}^{2}\,q_{m}^{2}}\left(q_{e}^{2}+q_{m}^{2}+r^{2}\right)\Bigg]\Big/4q_m^2.\label{LF2_BBb}
\end{align}
Taking into account the function \eqref{Sig1}, the metric function \eqref{aIV}, the scalar field \eqref{field_camp}, the potential \eqref{V_IV}, the Lagrangian \eqref{L2_BBb}, and the derivative of the Lagrangian \eqref{LF2_BBb}, we can now satisfy all equations \eqref{sol2}–\eqref{sol3} and \eqref{RC}, as well as all components of the equations of motion \eqref{EqF00}–\eqref{EqF22}.


\begin{figure*}[th!]
\centering
\includegraphics[scale=0.56]{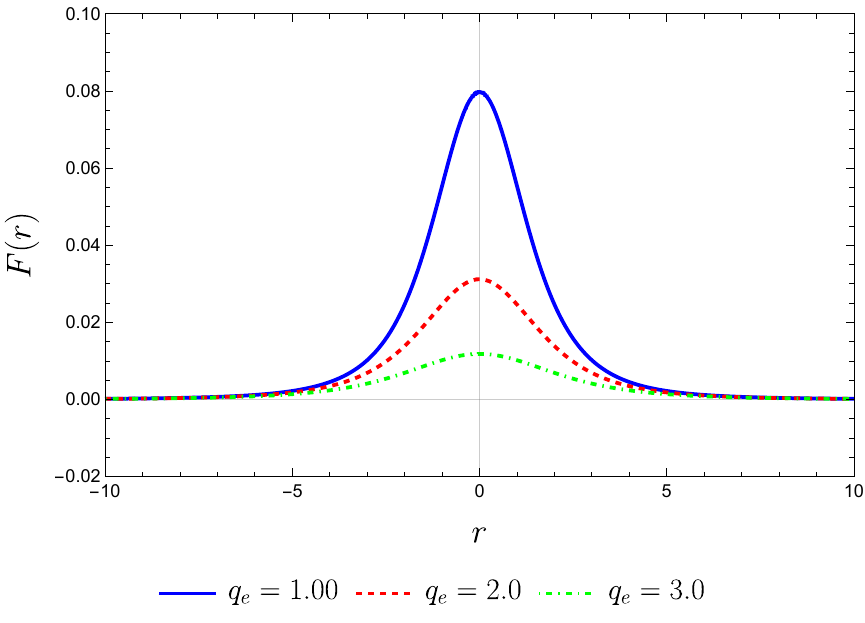}
	\hspace{0.75cm}
\includegraphics[scale=0.56]{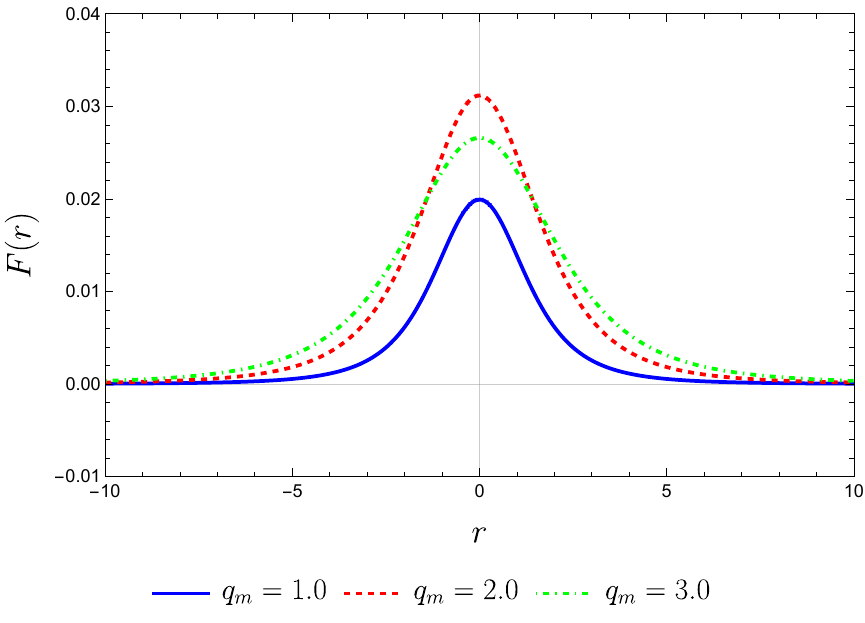}
\caption{The electromagnetic scalar $F$,  for different values of
the electric and magentic charges. In the left plot, we use the values of the constants as follows $\{M=4.0,\,q_m=2.0,\,c_1=1.0, \,c_2=1.0\}$. In the right plpot, we consider the values  $\{M=4.0,\,q_e=2.0,\,c_1=1.0, \,c_2=1.0\}$.} 
\label{fig_Fqe_IV}
\end{figure*}

We have also calculated the electromagnetic scalar for this solution and illustrate its behavior in Figs. \ref{fig_Fqe_IV}.
The left plot depicts three curves corresponding to different values of the electric charge ($q_e=1.0$, $q_e=2.0$ and $q_e=3.0$), similar as depicted in Fig. \ref{fig_Fqe_SV}. In the right plot, we analyze the variation of this electromagnetic scalar with the magnetic charge over three different curves ($q_m=1.0$, $q_m=2.0$ and $q_m=3.0$). 


By inverting the expression \eqref{field_camp}, we get $r(\cal{\varphi})$ and find
\begin{align}
& V(\varphi) = \frac{4M\left(q_{e}^{2}+q_{m}^{2}\right)}{5\kappa^{2}\left(\left(q_{e}^{2}+q_{m}^{2}\right)\text{sech}^{2}\left(\kappa\text{\ensuremath{\varphi}}\sqrt{\epsilon}\right)\right)^{5/2}}
\nonumber\\
&
+\frac{\cosh^{2}\left(\kappa\text{\ensuremath{\varphi}}\sqrt{\epsilon}\right)}{2\kappa^{2}}\left(4c_{1}-\frac{3c_{2}\tan\left(\text{\ensuremath{\varphi}}\sqrt{\kappa^{2}(-\epsilon)}\right)}{\left(q_{e}^{2}+q_{m}^{2}\right)^{3/2}}\right)
\nonumber\\
&
+\frac{c_{2}\tan^{-1}\left(\tan\left(\text{\ensuremath{\varphi}}\sqrt{\kappa^{2}(-\epsilon)}\right)\right)\left(\cosh\left(2\kappa\text{\ensuremath{\varphi}}\sqrt{\epsilon}\right)-2\right)}{2\kappa^{2}\left(q_{e}^{2}+q_{m}^{2}\right)^{3/2}}.\label{V3_BBb}
\end{align}

\begin{figure*}[th!]
\includegraphics[scale=0.56]{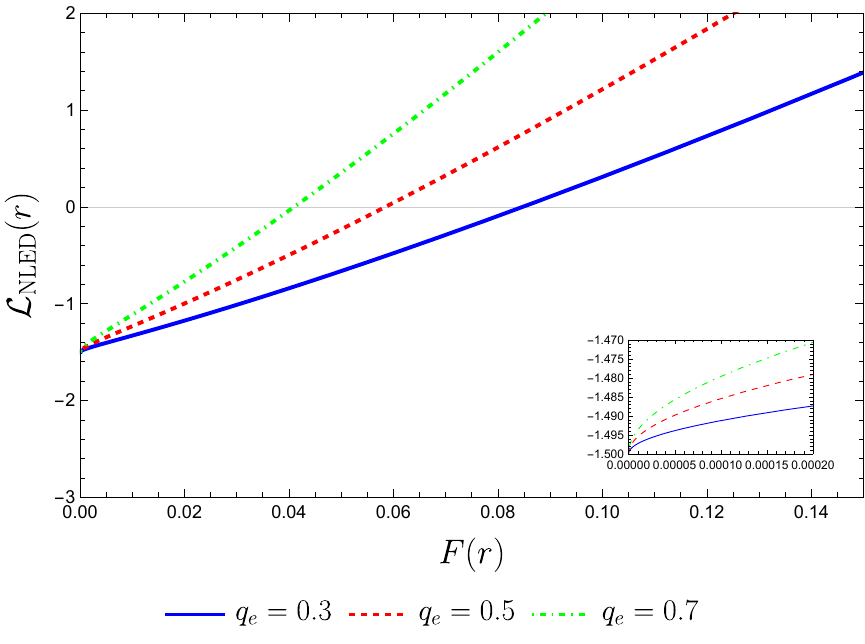}
	\hspace{0.75cm}
\includegraphics[scale=0.56]{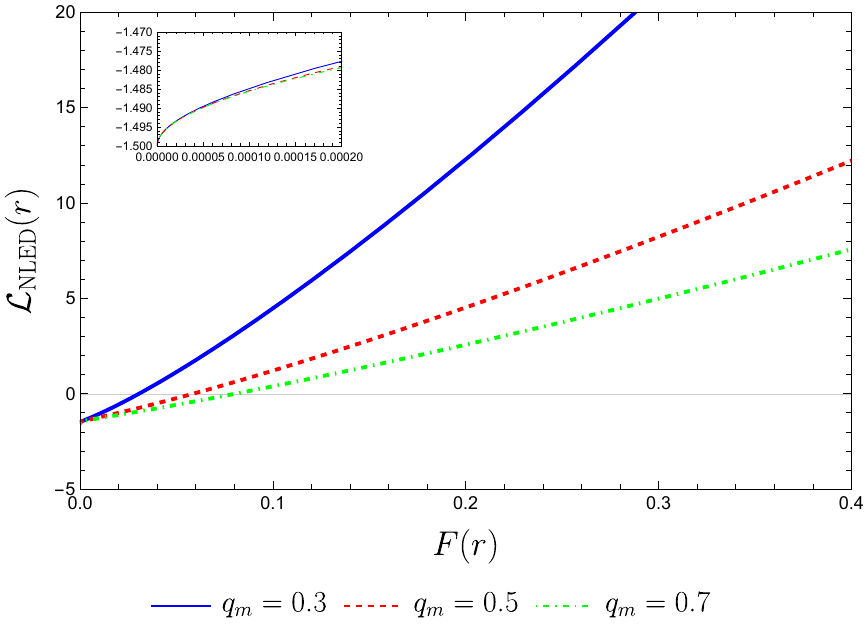}
\caption{The NLED Lagrangian ${\cal L}(F) $, for different values of the electric and magnetic charges. (i) In the left plot, we have used the values of the constants as follows $\{M=10.0,\,q_m=0.5,\,c_1=1.0, \,c_2=1.0\}$. (ii) In the right plot, we consider $\{M=10.0,\,q_e=0.5,\,c_1=1.0, \,c_2=1.0\}$.} 
\label{LxFqe_IV}
\end{figure*}

\begin{figure*}[ht!]
\includegraphics[scale=0.56]{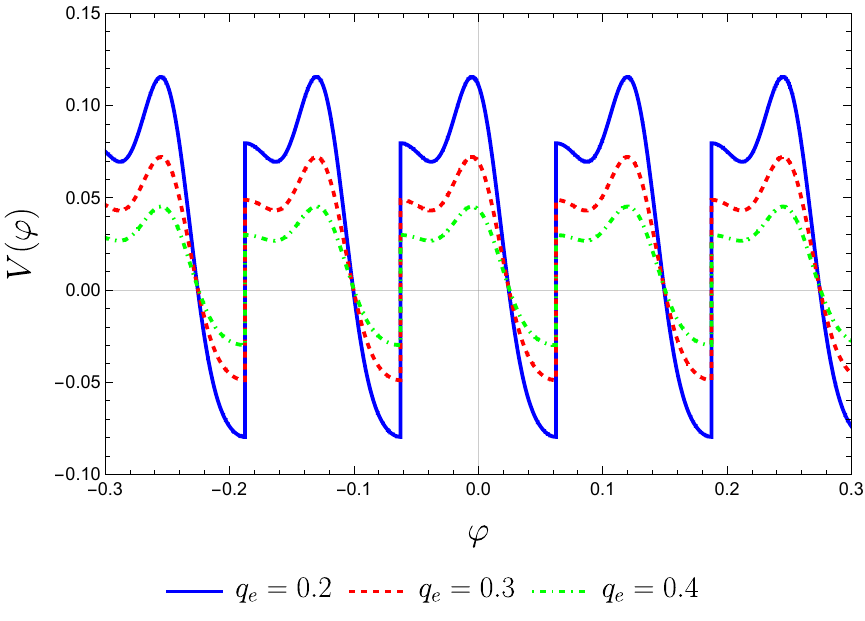}
	\hspace{0.75cm}
\includegraphics[scale=0.56]{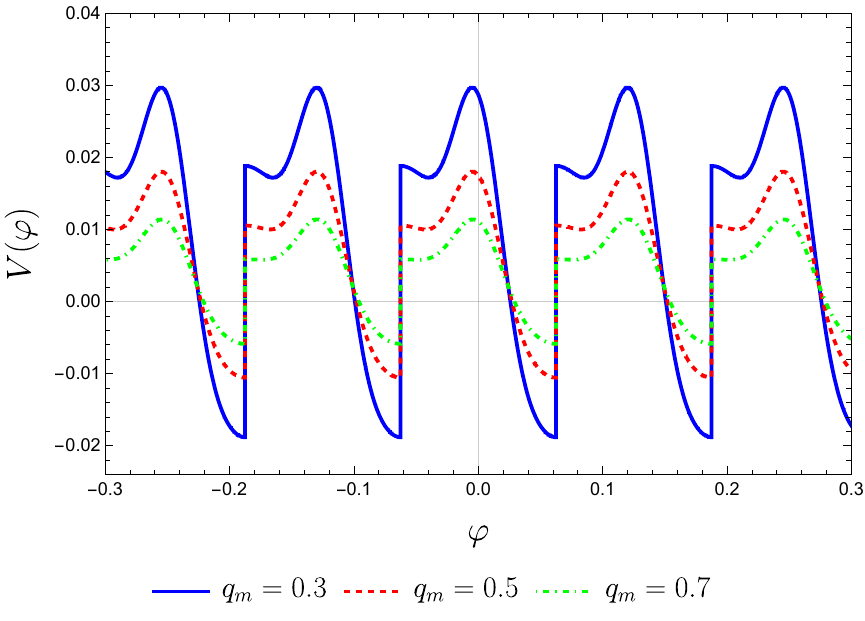}
 \includegraphics[scale=0.58]{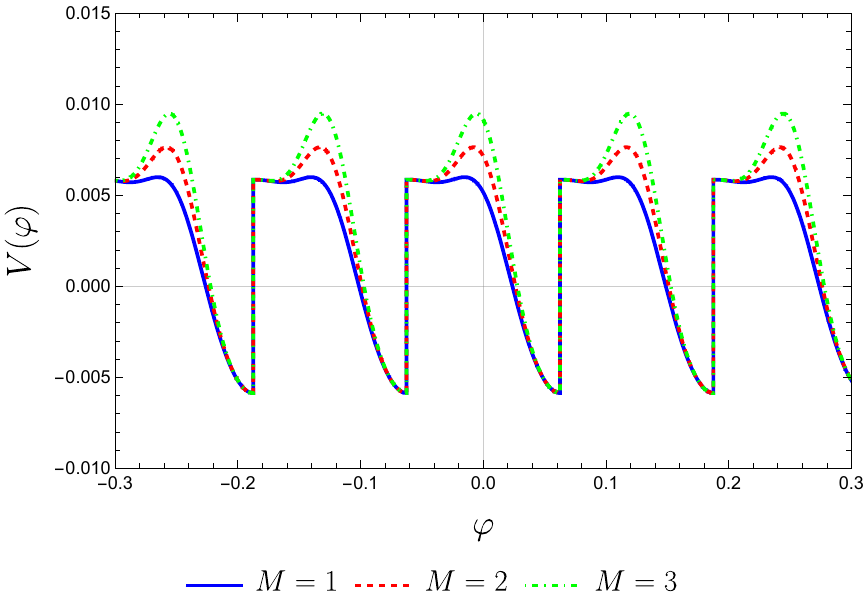}
\caption{The scalar field potential $V(\varphi)$, described by Eq.~\eqref{V3_BBb} taking $\epsilon=-1$. (i) In the left plot, we use the values $\{M=4.0,\,\,q_m=0.3,\,c_1=1.0,\,c_2=1.0\}$. (ii) In the rioght plot, we consider $\{M=4.0,\,\,q_e=0.5,\,c_1=1.0,\,c_2=1.0\}$. (iii) In the bellow plot, we consider $\{\,q_e=0.5,\,q_m=0.7,\,c_1=1.0,\,c_2=1.0\}$.} 
\label{Vxphi2_BB}
\end{figure*}

In Figs.\;\ref{LxFqe_IV}, we illustrate the behavior of the Lagrangian density of NLED as a function of $F$ using parametric graphs for different values of the electric charge ($q_e=0.3$, $q_e=0.5$ and $q_e=0.7$) and the magnetic charge ($q_m=0.3$, $q_m=0.5$ and $q_m=0.7$), respectively. In addition, we present the behavior of the scalar field potential given by Eq.~\eqref{V3_BBb} in three different diagrams, each showing three curves with variations of the electric charge ($q_e=0.2$, $q_e=0.3$ and $q_e=0.4$), the magnetic charge ($q_m=0.3$, $q_m=0.5$ and $q_m=0.7$) and the mass ($M=1.0$, $M=2.0$ and $M=3.0$), as shown in Figs. \ref{Vxphi2_BB}. 
We notice that the amplitude of the potential also varies when we change the values for electric charge, magnetic charge and mass.


\section{Fifth Dyon solution}\label{solV}
\subsection{Metric function and horizons}

In this solution, we also require that the Lagrangian described by Eq. \eqref{L_BB} is real. To achieve this, we model the argument root that arises from this Lagrangian according to the following condition:
\begin{eqnarray}
&& \Sigma(r)^{4}\left[\Sigma(r)^{2}A''(r)-2A(r)\left(\Sigma(r)\Sigma''(r)+\Sigma'(r)^{2}\right)+2\right]^{2} 
\nonumber
\\
&& \qquad
 -64q_{e}^{2}\,q_{m}^{2}=q_{e}^{2}\,q_{m}^{2}.
\end{eqnarray}
Again, after considering the function described by Eq. \eqref{Sig1}, we obtain the following simplified form of the above expression
\begin{eqnarray}
\left[A''(r)\left(q_{e}^{2}+q_{m}^{2}+r^{2}\right)-2A(r)+2\right]^{2} \times
\nonumber
\\
\times \left(q_{e}^{2}+q_{m}^{2}+r^{2}\right)^{2}=65q_{e}^{2}q_{m}^{2}\,, \label{eqA}
\end{eqnarray}
which yields the following solution for the metric function
\begin{align}
&A(r)=\Bigg\{\Big[4c_{1}\left(q_{e}^{2}+q_{m}^{2}\right)\left(q_{e}^{2}+q_{m}^{2}+r^{2}\right)+2r(c_{2}-2r) 
\nonumber\\
&
+\sqrt{65}\sqrt{q_{e}^{2}q_{m}^{2}}\Big]\times\left(q_{e}^{2}+q_{m}^{2}\right)+\tan^{-1}\left(\frac{r}{\sqrt{q_{e}^{2}+q_{m}^{2}}}\right)\times
\nonumber\\
&
\quad \times2\sqrt{q_{e}^{2}+q_{m}^{2}}\left[c_{2}\left(q_{e}^{2}+q_{m}^{2}+r^{2}\right)+\sqrt{65}r\sqrt{q_{e}^{2}q_{m}^{2}}\right]+
\nonumber\\
&
\sqrt{65}\sqrt{q_{e}^{2}\,q_{m}^{2}}
\left(q_{e}^{2}+q_{m}^{2}+r^{2}\right)
\left[\tan^{-1}\left(\frac{r}{\sqrt{q_{e}^{2}+q_{m}^{2}}}\right)\right]^2\Bigg\}
\nonumber\\
&
\quad \times\frac{1}{4\left(q_{e}^{2}+q_{m}^{2}\right)^{2}}\,
.\label{aV}
\end{align}

\begin{figure}[th!]
\includegraphics[scale=0.56]{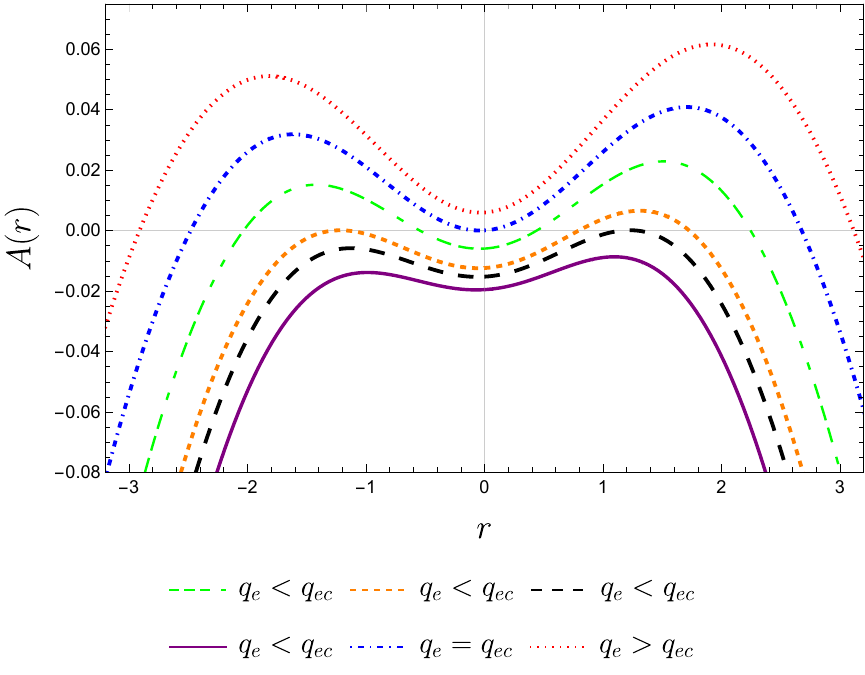}
S\caption{The  metric function $A(r) $, given by Eq.~\eqref{aV} for three electric and magnetic charge scenarios. We use the values of the constants as follows $\{\,q_m=2.0,\,c_1=-0.13, \,c_2=0.01\}$.} 
\label{figaqec}
\end{figure}

\begin{figure}[th!]
\includegraphics[scale=0.56]{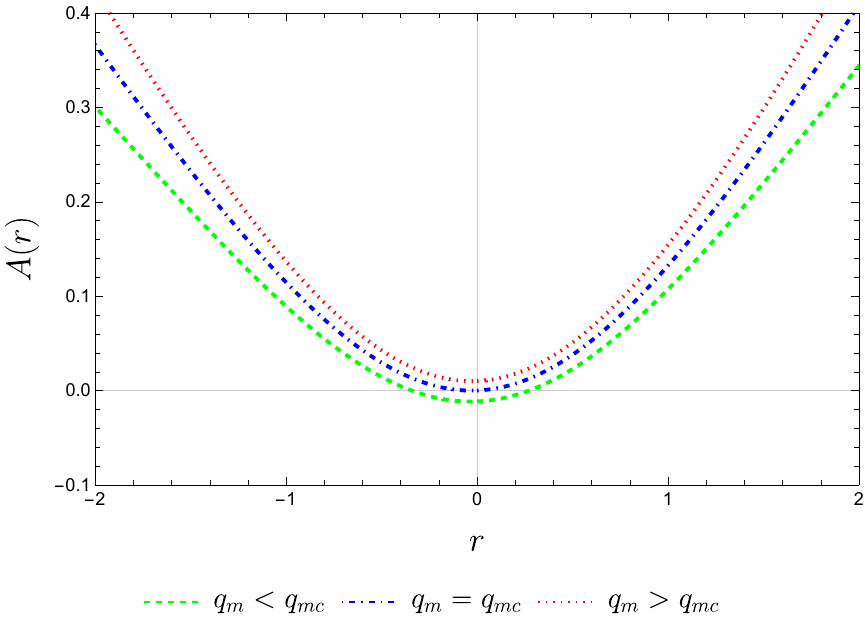}
S\caption{The  metric function $A(r) $, given by Eq.~\eqref{aV} for three electric and magnetic charge scenarios. Here, we consider $\{\,q_e=1.75,\,c_1=-0.205, \,c_2=0.03 \}$.} 
\label{figbqec}
\end{figure}

We solve Eqs.\;\eqref{rH} and \eqref{der_a} simultaneously for this model, determining the critical electric charge from the metric function \eqref{aV}. We obtained the value $q_{ec} = 0.619$, considering the constants defined as: $q_m = 2.0$, $c_1 = -0.13$, and $c_2 = 0.01$. In Fig. \ref{figaqec}, we present the behavior of the metric function \eqref{aV} in relation to the radial coordinate $r$ for three distinct scenarios of the electric charge: $q_e < q_{ec}$, $q_e = q_{ec}$, and $q_e > q_{ec}$.
When the charge is less than the critical charge, i.e., $q_e < q_{ec}$, we observe different behaviors. In the case described by the green curve, the formation of four horizons occurs. In the orange dashed curve, we see the formation of three horizons. In the black curve, there is only one horizon, while in the purple curve, no horizons are formed.
For the critical case, where $q_e = q_{ec}$, we identify the formation of up to three horizons. Finally, when the electric charge exceeds the critical value, i.e., $q_e > q_{ec}$, the formation of two horizons is observed.

For the critical magnetic charge, we obtained the value $q_m = 0.790$ by defining the constants as $q_e = 1.75$, $c_1 = -0.205$, and $c_2 = 0.03$. In Fig. \ref{figbqec}, we present the behavior of the metric function \eqref{aV} as a function of the radial coordinate $r$ for three different scenarios of the magnetic charge: $q_m > q_{mc}$, $q_m = q_{mc}$, and $q_m < q_{mc}$. In this case, we observe a similar horizon behavior to that of the cases presented in the previous sections.

Once again, we will ensure that the metric function \eqref{aV} is asymptotically flat. To achieve this, we will now consider the following expression for the constant $c_2$
\begin{align}
c_{2}=&\Big[-16\left[c_{1}q_{e}^{4}+q_{e}^{2}\left(2c_{1}q_{m}^{2}-1\right)+c_{1}q_{m}^{4}\right]+16q_{m}^{2}
\nonumber
\\
&
-\sqrt{65}\pi^{2}\sqrt{q_{e}^{2}\,q_{m}^{2}}\Big]\big/\left(4\pi\sqrt{q_{e}^{2}+q_{m}^{2}}\right). \label{c2_V}
\end{align}
With this choice, the metric function \eqref{aV} becomes asymptotically flat. Thus, by substituting the constant \eqref{c2_V} into the metric function \eqref{aV}, we obtain the following result
\begin{align}
&A(r)=\frac{1}{8\pi\left(q_{e}^{2}+q_{m}^{2}\right)^{2}}\Bigg\{\left(q_{e}^{2}+q_{m}^{2}\right)^{2}\left(q_{e}^{2}+q_{m}^{2}+r^{2}\right)\times
\nonumber
\\
& 
8\pi c_{1}-16c_1r\left(q_{e}^{2}+q_{m}^{2}\right)^{5/2}+\left(2\sqrt{q_{e}^{2}+q_{m}^{2}}-\pi r\right)\times
\nonumber
\\
& 
\times \sqrt{q_{e}^{2}+q_{m}^{2}} \left(8r\sqrt{q_{e}^{2}+q_{m}^{2}}+\pi\sqrt{65}\sqrt{q_{e}^{2}\,q_{m}^{2}}\right)
\nonumber
\\
& 
+2\sqrt{65}\pi\sqrt{q_{e}^{2}\,q_{m}^{2}}\left(q_{e}^{2}+q_{m}^{2}+r^{2}\right)\tan^{-1}\left(\frac{r}{\sqrt{q_{e}^{2}+q_{m}^{2}}}\right)^{2}
\nonumber
\\
& 
+\Bigg[4\sqrt{65}\pi r\sqrt{q_{e}^{2}\,q_{m}^{2}}\sqrt{q_{e}^{2}+q_{m}^{2}}+ \Big[16q_{m}^{2}
\nonumber
\\
& 
-16\left(c_{1}q_{e}^{4}+q_{e}^{2}\left(2c_{1}q_{m}^{2}-1\right)+c_{1}q_{m}^{4}\right)-\sqrt{65}\pi^{2}\sqrt{q_{e}^{2}\,q_{m}^{2}}\Big]
\nonumber
\\
& \times\left(q_{e}^{2}+q_{m}^{2}+r^{2}\right)\Bigg]\tan^{-1}\left(\frac{r}{\sqrt{q_{e}^{2}+q_{m}^{2}}}\right)\Bigg\}
.\label{aV2}
\end{align}
We also calculate the Komar mass, described by Eq. \eqref{m_Komar2}, for this metric function, Eq. \eqref{aV2}, whose explicit form is now
\begin{align}
M_{K}=
\frac{ {\bar c}_1}{24\pi},\label{Kom_A5}
\end{align}
where, we substitute 
\begin{align}
c_{1}=&\frac{\pi^{2}\sqrt{65}\sqrt{q_{e}^{2}q_{m}^{2}}+16q_{e}^{2}+16q_{m}^{2}-\bar{c}_{1}\sqrt{q_{e}^{2}+q_{m}^{2}}}{16\left(q_{e}^{2}+q_{m}^{2}\right)^{2}},
\end{align}
and $\bar{c}_1$ is a new positive real numerical constant independent of the charges $q_e$ and $q_m$, so that the Komar mass is also independent.

\begin{figure}[t!]
\includegraphics[scale=0.55]{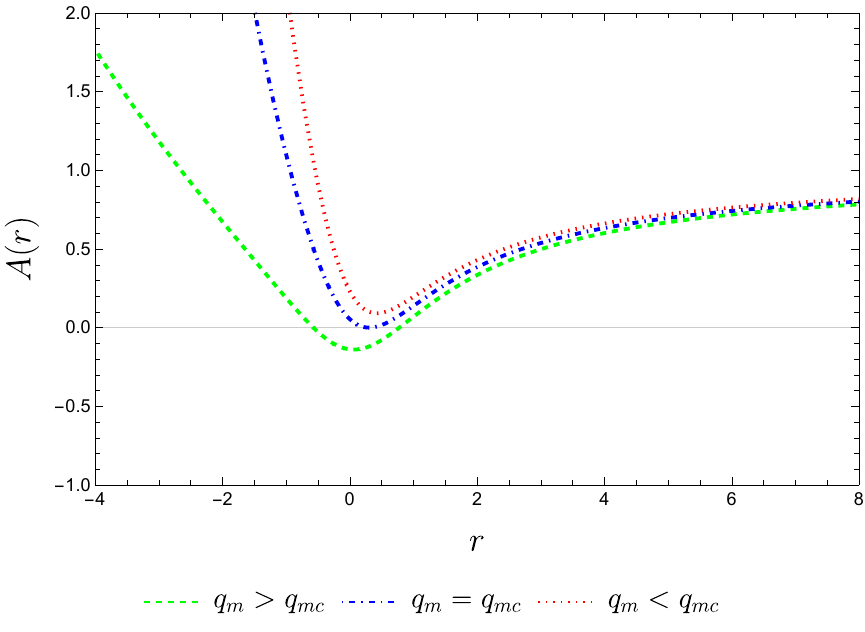}
\caption{The  metric function $A(r) $, given by Eq.~\eqref{aV2} for three magnetic charge scenarios. We have used the values of the constants as follows $\{q_e=0.5,\,c_1=-0.75\}$. } 
\label{figA2V}
\end{figure}

We also calculated the critical magnetic charge based on the metric function described by Eq. \eqref{aV2}, obtaining the value $q_m = 0.903$ by setting the constants to $q_e = 0.5$ and $c_1 = -0.75$. In Fig. \ref{figA2V}, we illustrate the behavior of the metric function \eqref{aV2} as a function of the radial coordinate $r$ for three different scenarios of the magnetic charge: $q_m > q_{mc}$, $q_m = q_{mc}$, and $q_m < q_{mc}$. We observe that this metric function exhibits asymmetric behavior: for large values of $r$, the function becomes constant in the region where $r > 0$, whereas in the region where $r < 0$, the function $A(r)$ increases indefinitely.

\subsection{Kretschmann scalar}

We also check the $r \rightarrow 0$ limit for the Kretschmann scalar of this solution, as presented below
\begin{align}
&\lim_{r\rightarrow0}K\left(r\right)=2c_{1}\left(6c_{1}+\frac{5\sqrt{65}\sqrt{q_{e}^{2}\,q_{m}^{2}}-4\left(q_{e}^{2}+q_{m}^{2}\right)}{\left(q_{e}^{2}+q_{m}^{2}\right)^{2}}\right)
\nonumber
\\
&
\qquad +\Bigg[715q_{m}^{2}\left(q_{e}^{2}+q_{m}^{2}\right)+32\left(q_{e}^{2}+q_{m}^{2}\right)^{2}-715q_{m}^{4}
\nonumber
\\
&
\qquad -24\sqrt{65}\sqrt{q_{e}^{2}\,q_{m}^{2}}\left(q_{e}^{2}+q_{m}^{2}\right)\Bigg]\Big/4\left(q_{e}^{2}+q_{m}^{2}\right)^{4}.
\end{align}
we observe that, in this limit, the Kretschmann scalar is regular. On the other hand, if we examine the $r \rightarrow \infty$ limit of the Kretschmann scalar, we obtain
\begin{align}
&\lim_{r\rightarrow\infty}K\left(r\right)=\frac{3}{32\left(q_{e}^{2}+q_{m}^{2}\right)^{3}}\Bigg[\sqrt{\frac{1}{q_{e}^{2}+q_{m}^{2}}}\bigg(16c_{1}\left(q_{e}^{4}+q_{m}^{4}\right)
\nonumber
\\
& 
+16q_{e}^{2}\left(2c_{1}q_{m}^{2}-1\right)+\pi^{2}\sqrt{65}\sqrt{q_{e}^{2}\,q_{m}^{2}}-16q_{m}^{2}\bigg)+4\pi c_{2}\Bigg]^2
,\label{K4}
\end{align}
and find that the Kretschmann scalar also remains regular in this case.

\begin{figure}[th!]
\includegraphics[scale=0.55]{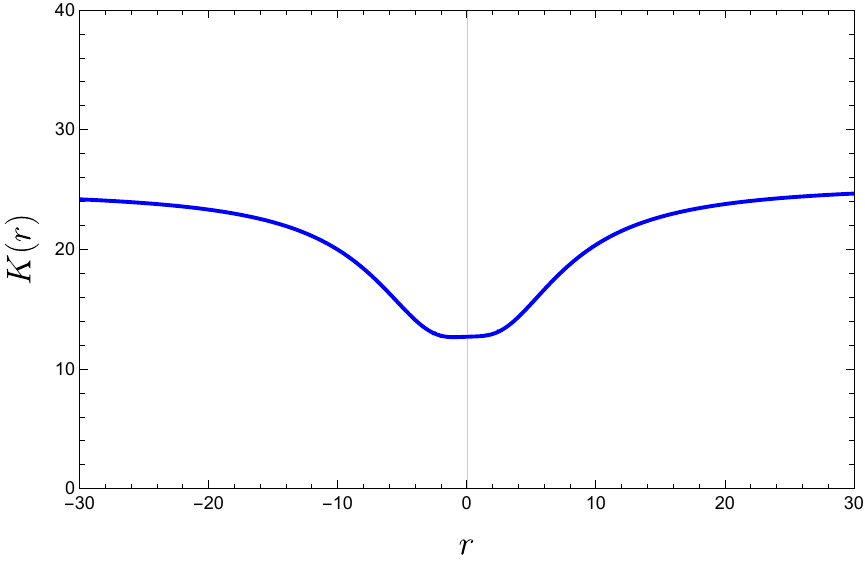}
\caption{The behavior of the Kretschmann scalar $K(r)$. We have used the values of the constants as follows $\{q_e=2\,\,q_m=5,\,c_1=c_2=1\}$ .} 
\label{fig_K_V}
\end{figure}

Although we do not present the explicit form of the Kretschmann scalar for this solution due to its cumbersome algebraic expression, we illustrate its behavior in Fig. \ref{fig_K_V}.

\subsection{Matter Lagrangian and Potential reconstruction}

Given the metric function obtained for this solution, Eq. \eqref{aV} and the function \eqref{Sig1},  after substituting into the solution \eqref{V}, we are able to determine the form of the potential in terms of the radial coordinate for this solution, as
\begin{align}
&V(r)=\Bigg\{3\sqrt{65}\sqrt{q_{e}^{2}q_{m}^{2}}\left[\left(q_{e}^{2}+q_{m}^{2}\right)^{2}-r^{4}\right]-2\left(q_{e}^{2}+q_{m}^{2}\right)
\nonumber\\
& \qquad
\times \Big[ 8\left(q_{e}^{2}+q_{m}^{2}\right)^{2}\left[c_{1}\left(q_{e}^{2}+q_{m}^{2}\right)-1\right]+\sqrt{65}q_{e}^{2}\sqrt{q_{e}^{2}\,q_{m}^{2}}
\nonumber\\
&
+\sqrt{65}q_{m}^{2}\sqrt{q_{e}^{2}\,q_{m}^{2}}-6c_{2}r\left(q_{e}^{2}+q_{m}^{2}\right)-6c_2 r^3+8r^2
\nonumber\\
&
\qquad \times \left(q_{e}^{2}+q_{m}^{2}\right)\left[c_{1}\left(q_{e}^{2}+q_{m}^{2}\right)-1\right] \Big] +4\left(q_{e}^{2}+q_{m}^{2}+r^{2}\right)
\nonumber\\
& \qquad
\times \sqrt{q_{e}^{2}+q_{m}^{2}} \left[c_{2}\left(q_{e}^{2}+q_{m}^{2}+3r^{2}\right)+3\sqrt{65}r\sqrt{q_{e}^{2}\,q_{m}^{2}}\right]\,
\nonumber\\
& \quad
\times \tan^{-1}\left(\frac{r}{\sqrt{q_{e}^{2}+q_{m}^{2}}}\right)+2\sqrt{65}\sqrt{q_{e}^{2}\,q_{m}^{2}}\left(q_{e}^{2}+q_{m}^{2}+r^{2}\right)
\nonumber\\
&
\quad\times \left(q_{e}^{2}+q_{m}^{2}+3r^{2}\right)\tan^{-1}\left(\frac{r}{\sqrt{q_{e}^{2}+q_{m}^{2}}}\right)^{2}\Bigg\}
\nonumber\\
& \quad
\times \left(-\frac{1}{8\kappa^{2}\left(q_{e}^{2}+q_{m}^{2}\right)^{2}\left(q_{e}^{2}+q_{m}^{2}+r^{2}\right)^{2}}\right)
.\label{V_BBb}
\end{align}

For this model, we then have the following electromagnetic quantities
\begin{align}
{\cal L}_{\textrm{NLED}}(r) =&-\frac{3}{2}c_{1}+\frac{\sqrt{q_{e}^{2}\,q_{m}^{2}}}{8\left(q_{e}^{2}+q_{m}^{2}+r^{2}\right)^{2}}+\frac{3}{2\left(q_{e}^{2}+q_{m}^{2}\right)}
\nonumber\\
&
-\frac{3\sqrt{65}\sqrt{q_{e}^{2}\,q_{m}^{2}}}{16\left(q_{e}^{2}+q_{m}^{2}\right)^{2}},
\label{L4_BB} 
\end{align}
and its derivative ${\cal L}_F$ Eq. \eqref{LF_BB}
\begin{align}
&{\cal L}_F(r) =\frac{\left(\sqrt{65}+1\right)\sqrt{q_{e}^{2}\,q_{m}^{2}}}{8q_{m}^{2}}.\label{LF4_BB}
\end{align}
Taking into account the function \eqref{Sig1}, the metric function \eqref{aV},  the scalar field \eqref{field_camp}, the potencial \eqref{V_BBb}, the Lagrangean \eqref{L4_BB} and the derivative of the Lagrangean \eqref{LF4_BB}, these satisfy all of equations \eqref{sol2}-\eqref{sol3} and \eqref{RC}, and all components of the equations of motion \eqref{EqF00}-\eqref{EqF22}.
Note that the derivative of the Lagrangian we obtain, Eq. \eqref{LF4_BB}, depends only on constant quantities. 

The electromagnetic scalar for this solution has the form of:
\begin{align}
    F(r)=\frac{q_{m}^{2}}{\left(\sqrt{65}+1\right)\left(q_{e}^{2}+q_{m}^{2}+r^{2}\right)^{2}}.\label{FV}
\end{align}

\begin{figure*}[th!]
\centering
\includegraphics[scale=0.55]{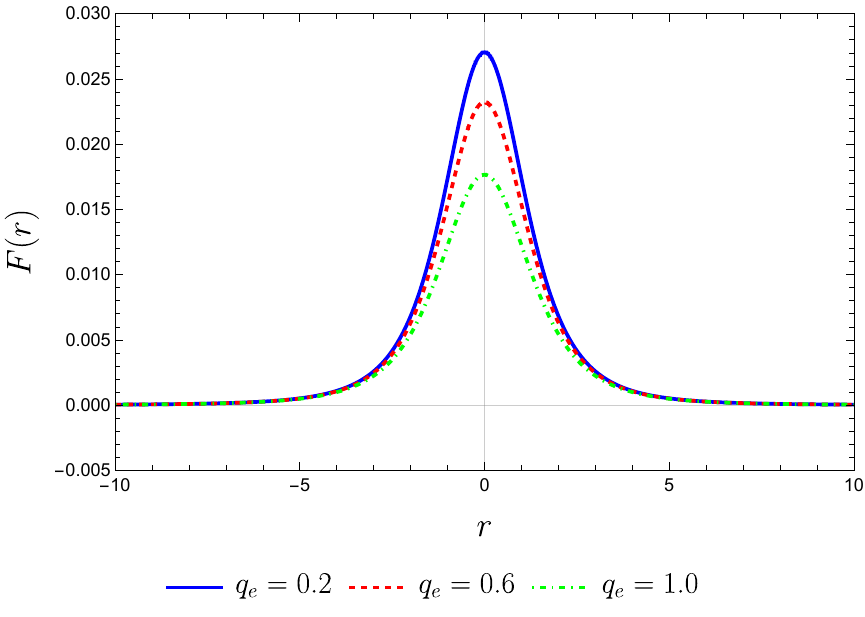}
	\hspace{0.75cm}
\includegraphics[scale=0.55]{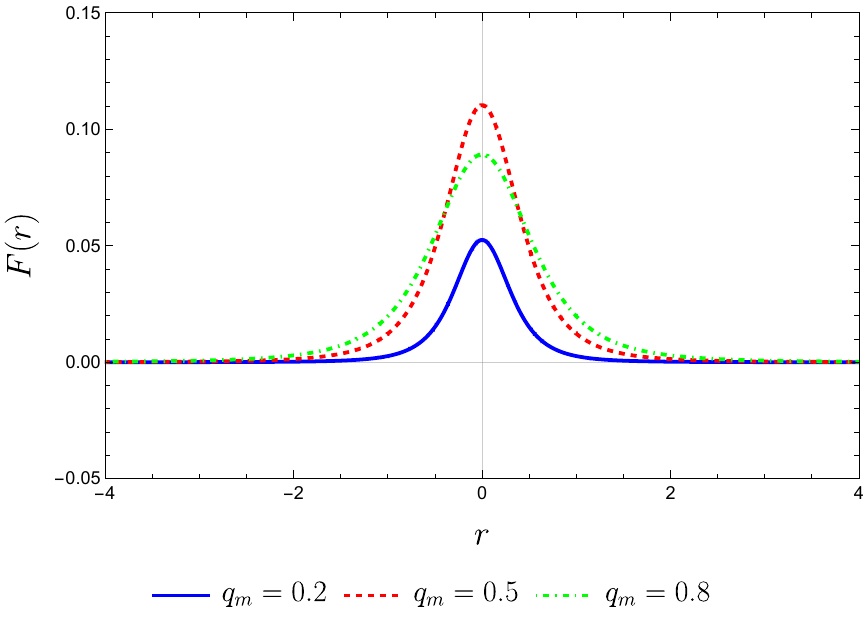}
\caption{The electromagnetic scalar $F$, given by Eq.~\eqref{FV}, for different values of
the electric and magnetic charges. (i) In the left plot, we have used the values of the constants as follows $\{q_m=2.0,\,c_1=1.0, \,c_2=1.0\}$. In the right plot, we consider the following values $\{q_e=0.5,\,c_1=1.0, \,c_2=1.0\}$.} 
\label{fig_Fqe_V}
\end{figure*}

We illustrate the behavior of the electromagnetic scalar \eqref{FV} in Figs. \ref{fig_Fqe_V}.
In the first figure, we plot three curves corresponding to different values of the electric charge ($q_e=0.2$, $q_e=0.6$ and $q_e=1.0$), as depicted in Fig. \ref{fig_Fqe_V}. In the second figure, we analyze the variation of this electromagnetic scalar with the magnetic charge over three different curves ($q_m=0.2$, $q_m=0.5$ and $q_m=0.8$), as shown in the right plot of Fig. \ref{fig_Fqe_V}.

On the other hand, inverting  the expressions above to obtain  $r(F)$ and $r(\cal{\varphi})$, we determine
\begin{align}
   {\cal L} _{\rm NLED} (F) =&	-\frac{3}{2}c_{1}+\left(\sqrt{65}+1\right)\frac{\sqrt{q_{e}^{2}\,q_{m}^{2}}}{8q_{m}^{2}}F
   \nonumber\\
&
+\frac{3\left(8\left(q_{e}^{2}+q_{m}^{2}\right)-\sqrt{65}\sqrt{q_{e}^{2}\,q_{m}^{2}}\right)}{16\left(q_{e}^{2}+q_{m}^{2}\right)^{2}}.
 \label{L_V} 
\end{align}   
and
\begin{align}
& V(\varphi) = \Bigg\{\frac{24c_{2}\sqrt{\epsilon}\sqrt{q_{e}^{2}+q_{m}^{2}}\sinh\left(2\kappa\text{\ensuremath{\varphi}}\sqrt{\epsilon}\right)}{\sqrt{(-\epsilon)}}-8\sqrt{65}\times
\nonumber\\
&
\times \sqrt{q_{e}^{2}\,q_{m}^{2}}\left[\cosh\left(2\kappa\text{\ensuremath{\varphi}}\sqrt{\epsilon}\right)-2\right]\tanh^{-1}\left(\tanh\left(\kappa\text{\ensuremath{\varphi}}\sqrt{\epsilon}\right)\right)^{2}
\nonumber\\
&
+\sqrt{65}\sqrt{q_{e}^{2}\,q_{m}^{2}}\left[\cosh\left(4\kappa\text{\ensuremath{\varphi}}\sqrt{\epsilon}\right)-8\cosh\left(2\kappa\text{\ensuremath{\varphi}}\sqrt{\epsilon}\right)\right]
\nonumber\\
&
+32\left(q_{e}^{2}+q_{m}^{2}\right)\left[c_{1}\left(q_{e}^{2}+q_{m}^{2}\right)-1\right]\left[1+\cosh\left(2\kappa\text{\ensuremath{\varphi}}\sqrt{\epsilon}\right)\right]
\nonumber\\
&
+\frac{8\tanh^{-1}\left[\tanh\left(\kappa\text{\ensuremath{\varphi}}\sqrt{\epsilon}\right)\right]}{\kappa\sqrt{\epsilon}}\Bigg[ -4c_{2}\sqrt{q_{e}^{2}+q_{m}^{2}}\sqrt{\kappa^{2}(-\epsilon)}
\nonumber\\
&
+2c_{2}\sqrt{q_{e}^{2}+q_{m}^{2}}\sqrt{\kappa^{2}(-\epsilon)}\cosh\left(2\kappa\text{\ensuremath{\varphi}}\sqrt{\epsilon}\right)+3\sqrt{65}\kappa\sqrt{\epsilon}\times
\nonumber\\
&
\times
\sqrt{q_{e}^{2}\,q_{m}^{2}}\sinh\left(2\kappa\text{\ensuremath{\varphi}}\sqrt{\epsilon}\right)\Bigg]
+3\sqrt{65}\sqrt{q_{e}^{2}\,q_{m}^{2}}\Bigg\}\times
\nonumber\\
&
\times\frac{1}{32\kappa^{2}\left(q_{e}^{2}+q_{m}^{2}\right)^{2}}
.\label{V_sol_V}
\end{align}

These expressions, Eqs.~\eqref{L_V} and \eqref{V_sol_V}, represent the material content of this solution. We have obtained a result that was expected to be unequivocal for black-bounce solutions. Specifically, for this class of solutions, when coupling the equations of motion of a given theory with the material content described by NLED and a scalar field, a Lagrangian that is analytically linear, such as the one obtained in this solution and given by Eq.~\eqref{L_V}, had never been found before. Previously, only analytical nonlinear Lagrangians had been obtained.

\begin{figure*}[htb!]
\includegraphics[scale=0.55]{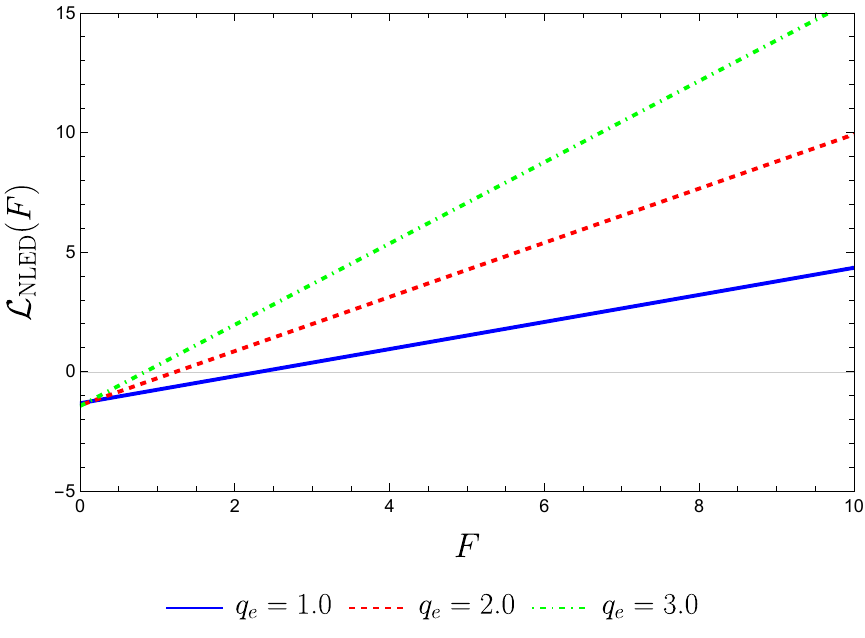}
	\hspace{0.75cm}
\includegraphics[scale=0.55]{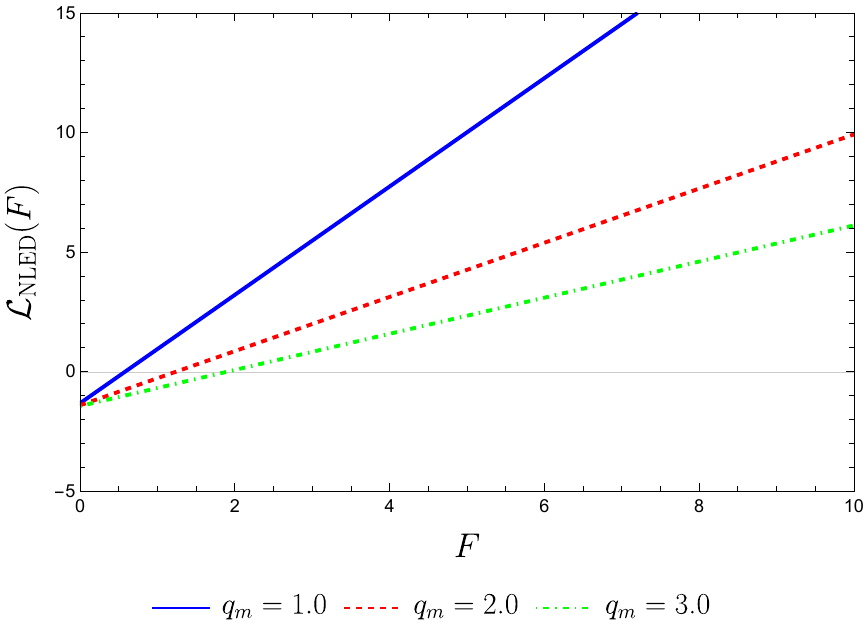}
\caption{The NLED Lagrangian ${\cal L}(F) $, given by Eq.~\eqref{L_V} for different values of the electric and magnetic charges. In the left plot, we considered $\{q_e=1.0,\,q_e=2.0,\,q_e=3.0\}$, and used the values of the constants as follows $\{q_m=2.0,c_1=1.0\}$. In the right plot, we consider $\{q_m=1.0,\,q_m=2.0,\,q_m=3.0\}$, and used the values of the constants as follows $\{q_e=2.0,c_1=1.0\}$.  } 
\label{LxFqe_V}
\end{figure*}

\begin{figure*}[htb!]
\includegraphics[scale=0.55]{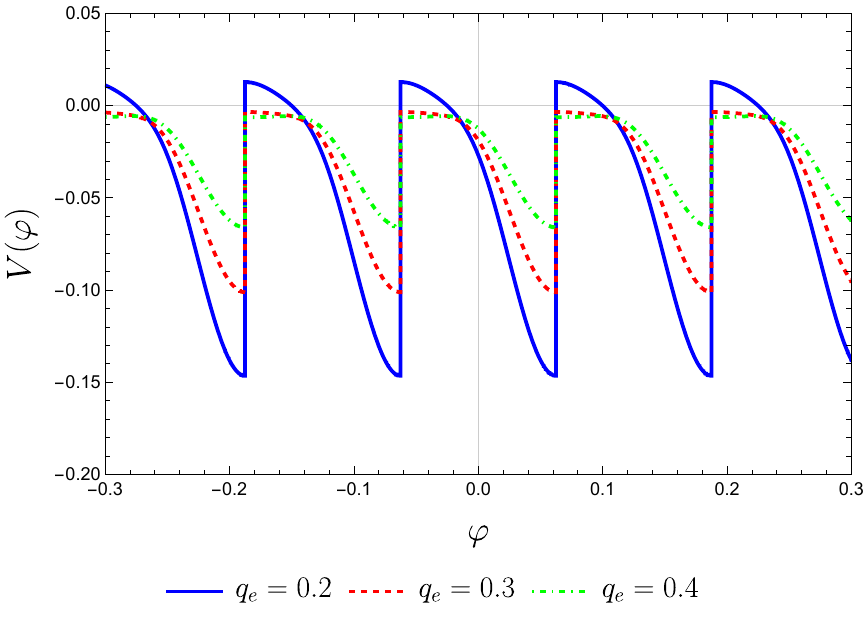}
	\hspace{0.75cm}
\includegraphics[scale=0.55]{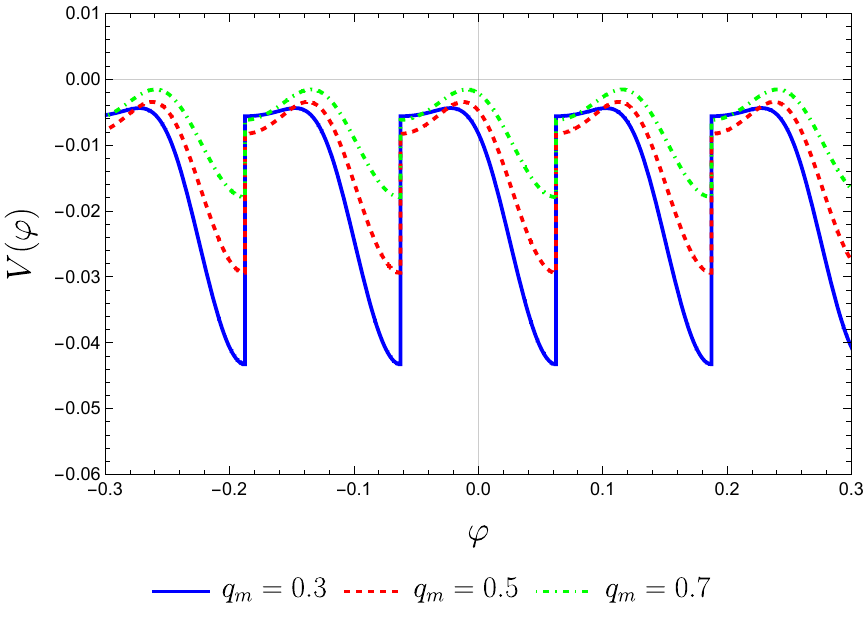}
\caption{The scalar field potential $V(\varphi)$, described by Eq.~\eqref{V_sol_V} for different values of the electric and magnetic charges, taking into account $\epsilon=-1$. (i) In the left plot, we consider $\{q_e=0.2,\,q_e=0.3,\,q_e=0.4\}$, for the values $\{q_m=0.3,\,c_1=1.0,\,c_2=1.0\}$. (ii) In the right plot, we consider $\{q_m=0.3,\,q_m=0.5,\,q_m=0.7\}$, for the values $\{q_e=0.5,\,c_1=1.0,\,c_2=1.0\}$.} 
\label{fig_Vxphi_V}
\end{figure*}

In Figs.\;\ref{LxFqe_V}, we illustrate the behavior of the Lagrangian density of NLED as a function of $F$, Eq. \eqref{L_V}, using plots for different values of the electric charge ($q_e=1.0$, $q_e=2.0$ and $q_e=3.0$) and the magnetic charge ($q_m=1.0$, $q_m=2.0$ and $q_m=3.0$), respectively. We present the behavior of the scalar field potential given by Eq.~\eqref{V_sol_V} in three different graphs, each showing three curves with variations of the electric charge ($q_e=0.2$, $q_e=0.3$ and $q_e=0.4$) and the magnetic charge ($q_m=0.3$, $q_m=0.5$ and $q_m=0.7$), as shown in Figs. \ref{fig_Vxphi_V}. We notice that the amplitude of the potential also varies when we modify the values for electric charge, magnetic charge and mass.

\section{Summary and Conclusion}\label{sec:concl}

In this article, we delved into the study of black-bounce solutions within the context of GR, focusing on static and spherically symmetric geometries. Our primary aim was to explore these solutions by coupling the GR field equations to a combination of NLED and a scalar field, which together act as the source of matter. This approach represents a novel departure from previous studies, which typically consider only magnetic charge or electric charge individually. By incorporating both magnetic and electric charges, we obtain what are known as dyon solutions, offering a more comprehensive view of these intriguing spacetime configurations.
We began by presenting the dyon solution within the well-established Reissner-Nordström model, where we do so without including the scalar field as a matter source. In this initial framework, we also calculated the Komar mass associated with this solution, a crucial quantity that we similarly computed for all the subsequent solutions developed.
Following this, we expanded on the BB solutions within the dyon context. These solutions introduced a regularization mechanism that avoids the central singularity traditionally present in black hole models, instead featuring a smooth ``bounce'' at the center. By developing these BB solutions in the presence of both electric and magnetic charges, we offered new insights into the nature of black bounce geometries and their potential to resolve some of the longstanding issues in gravitational physics, particularly regarding singularities and the structure of black holes.

In the first black-bounce model, we explored the solutions by generalizing the Simpson-Visser framework. In this approach, we focused on studying the horizons through numerical solutions, obtained by simultaneously solving the equations \eqref{rH} and \eqref{der_a}. This allowed us to determine the critical parameters: $q_{ec}$, $q_{mc}$, and $M_{c}$.
First, we analyzed the critical charges $q_{ec}$ and $q_{mc}$. We observed similar behaviors with respect to the horizons: 
\begin{itemize}
	\item No horizon exists for $q_e > q_{ec}$ or $q_m > q_{mc}$.
	\item When $q_e = q_{ec}$ and $q_m = q_{mc}$, the horizons become degenerate.
	\item For $q_e < q_{ec}$ and $q_m < q_{mc}$, two horizons appear.
\end{itemize}
Next, we determined the critical mass $M_c$. For $M > M_c$, a single event horizon is present on both sides of the bounce at $r = 0$, which merges into a degenerate horizon when $M = M_c$. In contrast, if $M < M_c$, a traversable wormhole forms. We also calculated the Komar mass and found that the metric parameter in this solution coincides with the Komar mass.
To further assess the model, we analyzed the Kretschmann scalar, which reveals regularity throughout the entire spacetime. Finally, through direct calculation, we verified that these black-bounce geometries are supported by a specific combination of matter contributions.

In the solutions discussed in Sections \ref{solIII} and \ref{solV}, we determined the metric function from the Lagrangian expression \eqref{L_BB} by imposing the condition that the term under the square root in this expression is a positive real number. Using the resulting metric functions, given by Eqs.\;\eqref{aIII} and \eqref{aV}, we investigated the possibility of event horizons' existence, depending on the choice of the solution parameters. The behaviors observed were analogous to those found in the Simpson-Visser dyon model.
A notable feature that emerged, which was already subtly present in the Simpson-Visser model for our dyon solution, is the evidence of asymmetry in the behavior of the metric functions. This asymmetry is clearly illustrated when comparing the regions $r > 0$ and $r < 0$, particularly through the event horizon curves for the two models.
Additionally, we found that, in general, the metric functions of these models, Eqs.\;\eqref{aIII} and \eqref{aV}, are not asymptotically flat. To achieve asymptotic flatness, we introduced the constants \eqref{c2_III} and \eqref{c2_V}. These constants enabled us to accurately determine the mass or gravitational energy of the system. Consequently, we derived the new metric functions, given by Eqs.\;\eqref{aIII2} and \eqref{aV2}, for which we also computed the Komar mass.
Finally, as was done in the Simpson-Visser dyon model, we verified the regularity of the Kretschmann scalar in both solutions, confirming its regular behavior throughout the entire spacetime. Furthermore, we determined the explicit forms of the NLED Lagrangian density and the scalar potential that configure these models.

In Section \ref{solIV}, we generalized the metric function described by Eq. \eqref{aIII} to incorporate mass into its configuration, as specified in Eq. \eqref{aIV}. From this newly obtained metric function, Eq. \eqref{aIV}, we investigated the possibility of horizon formation, discovering that the solutions exhibit behavior analogous to that found in the Simpson-Visser dyon model. 
A key observation is that this metric function also displays asymmetry in its behavior. As with the previous models, we made the metric function \eqref{aIV} asymptotically flat by choosing the constant $c_2$ appropriately, as described in Eq. \eqref{c2_V}. This choice allows us to determine the new metric function, as expressed in Eq. \eqref{aIII2}, for which we then computed the Komar mass.
Finally, we verified the regularity of the Kretschmann scalar for this solution and identified the explicit forms of the NLED Lagrangian density and the scalar potential that structure this model.

To conclude our discussion, we highlight that the model described by the metric function in Eq.\;\eqref{aV} offers a novel result in the literature on black bounce solutions. For the first time, a black bounce solution has been derived by coupling a linear Lagrangian from the electromagnetic sector with a scalar field possessing a potential, serving as the matter source. 
This innovative approach marks a significant step forward in the exploration of non-singular spacetimes.
In particular, we observe that the analytical expression for the Lagrangian $\mathcal{L}(F)$ in Eq. \eqref{L_V} is inherently nonlinear, which sets it apart from previous studies that only considered magnetic or electric charges in isolation.
Moreover, this previously unexplored aspect of the model presents exciting opportunities for further research. It paves the way for the study of black bounce solutions within the context of dyonic solutions in other modified theories of gravity, offering a promising path toward the development of new, more comprehensive frameworks for understanding gravitational phenomena beyond the singularities traditionally associated with black holes.

Finally, we can provide a physical interpretation of the dyonic configuration to derive some phenomenological implications with respect to geodesic motion. Next, we will briefly discuss an analysis of this motion in an dyonic context and make a comparison with isolated cases containing only electric or only magnetic charge.

For this analysis, we consider a hypothetical laboratory near a black hole, more specifically near an accretion disk. The ionization of the gas in the accretion disk occurs mainly due to the extremely high temperatures generated by the intense physical processes in this extreme environment. As the gas spirals towards the black hole, the friction between the different layers of the disk converts gravitational energy into heat, significantly increasing the temperature of the plasma. If this temperature exceeds several thousand or million Kelvin, the atoms are stripped of their electrons and a highly ionized plasma is formed. In addition, the emission of intense electromagnetic radiation —  especially high-energy photons — contributes to this process by colliding with the atoms and promoting their ionization.

Therefore, this environment provides ideal conditions for studying the motion of charged particles near the black hole, regardless of whether they are electrically or magnetically charged or carry both charges simultaneously. Analyzing the geodesic motion of such particles allows us to evaluate the phenomenological effects associated with the simultaneous presence of gravitational, electric and magnetic fields.

To investigate these differences, we analyze the geodesics associated with a dionic configuration. We start with the Euler-Lagrange equation,
\begin{equation}
    \frac{d}{d\tau}\left(\frac{\partial{\cal L}}{\partial\dot{x}^{\mu}}\right)=\frac{\partial{\cal L}}{\partial x^{\mu}},\label{EL}
\end{equation}
where $\tau$ stands for the affine parameter, the superscript ($\dot{}$) for the derivative to $\tau$ and the four-velocity is $\dot{x}^{\mu}$. 

We consider the following relativistic Lagrangian for a particle of mass $m$ and charge $Q$ is
\begin{equation}
    {\cal L} = \frac{1}{2}m g_{\mu\nu}\dot{x}^\mu\dot{x}^\nu+Q A_\mu \dot{x}^\mu.\label{L}
\end{equation}
We also use the constants of motion associated with the temporal and axial symmetries of the metric, corresponding to the energy $\varepsilon$ and the angular momentum $L$, which are obtained from the definition of the canonical moment 

\begin{equation}
    \pi^\mu=\frac{dx^\mu}{d\tau}+Q A^\mu ,
\end{equation}
so
\begin{equation}
    \varepsilon=\pi_t, \quad\quad L=- \pi_{\phi}.
\end{equation}

The general equation of motion, obtained from the Euler-Lagrange equations, takes the form
\begin{equation}
    \frac{d^{2}x^{\mu}}{d\tau^{2}}+\Gamma_{\phantom{\mu}\alpha\beta}^{\mu}\frac{dx^{\alpha}}{d\tau}\frac{dx^{\beta}}{d\tau}=\frac{Q}{m}F_{\phantom{\mu}\nu}^{\mu}\frac{dx^{\nu}}{d\tau}.
\end{equation}
where $F_{\phantom{\mu}\nu}^{\mu}$ is the electromagnetic field tensor \eqref{MaxFar}. 

Given the metric function \eqref{m}, the equations of motion  resulting from \eqref{EL} for the Lagrangian \eqref{L} are
\begin{align}
 &   \dot{t}	=\frac{1}{mA\left(r\right)}\left(\varepsilon+Qq_{e}\int\frac{1}{\Sigma^{4}\left(r\right){\cal L}_{F}\left(r\right)}dr\right)
 \nonumber
 \\
 &
\ddot{r}	=\frac{A^{\prime}\left(r\right)}{2A^{2}\left(r\right)}\left[\frac{1}{m^{2}}\left(\varepsilon+Qq_{e}\int\frac{1}{\Sigma^{4}\left(r\right){\cal L}_{F}\left(r\right)}dr\right)^{2}+\dot{r}^{2}\right]
\nonumber
 \\
 &
 -\frac{A\left(r\right)\Sigma\left(r\right)\Sigma^{\prime}\left(r\right)}{m^{2}}\left[\dot{\theta}^{2}+\frac{1}{\Sigma^{4}\left(r\right)\sin^{2}\theta}\left(L+Qq_{m}\cos\theta\right)^{2}\right]
\nonumber
 \\
 &
 -\frac{Q^{2}q_{e}}{m^{2}\Sigma^{4}\left(r\right){\cal L}_{F}\left(r\right)}\left(\varepsilon+Qq_{e}\int\frac{1}{\Sigma^{4}\left(r\right){\cal L}_{F}\left(r\right)}dr\right),
 \nonumber
 \\
 &
\ddot{\theta}	=-2\frac{\Sigma^{\prime}\left(r\right)}{\Sigma\left(r\right)}\dot{r}\dot{\theta}+\frac{\cos\theta}{m^{2}\Sigma^{4}\left(r\right)\sin^{3}\theta}\left(L+Qq_{m}\cos\theta\right)^{2}
\nonumber
 \\
 &
 -\frac{Qq_{m}}{m^{2}\Sigma^{4}\left(r\right)\sin\theta}\left(L+Qq_{m}\cos\theta\right),
    \nonumber
 \\
 &
\dot{\phi}	=\frac{1}{m\Sigma^{2}\left(r\right)\sin^{2}\theta}\left(L+Qq_{m}\cos\theta\right).
\end{align}

A detailed analysis of the equations of motion shows that the trajectory of a test particle is highly sensitive to the nature of its charge. Purely electric cases ($q_e\neq0$, $q_m=0$) or purely magnetic cases ($q_e=0$, $q_m\neq0$) already introduce significant modifications to the geodesics compared to the neutral case, by altering specific terms in the equations for $\dot{t}$, $\ddot{r}$, $\ddot{\theta}$, and $\dot{\phi}$.

However, the dyonic situation ($q_e\neq0$, $q_m\neq0$) represents a fundamentally different and more complex dynamical scenario. In this case, the equations of motion simultaneously incorporate terms dependent on the electric charge $q_e$  and terms dependent on the magnetic charge $q_m$. This coexistence, together with the implicit interaction between electric and magnetic effects, results in a unique orbital dynamics for the dyon, which cannot be simply described as a superposition of purely electric or purely magnetic cases. Dyonic trajectories therefore exhibit their own distinctive characteristics, governed by the intricate interplay between gravitational and electromagnetic fields, differing markedly from the trajectories observed in scenarios with only one type of charge.

Dyonic solutions for black holes are motivated by both theoretical interest and potential phenomenological and experimental implications. Although astrophysical black holes are thought to be practically neutral due to rapid charge neutralization by the surrounding plasma, the study of configurations that simultaneously possess electric and magnetic charges enables the investigation of how such charges affect the causal structure, particle dynamics, and observable phenomena such as gravitational lensing and black hole shadows.

The simultaneous presence of electric and magnetic charges directly alters the metric of spacetime and affects key properties such as the position of event horizons and Cauchy horizons as well as the effective potential of particles and photons. These changes influence, for example, the position and stability of orbits, the radius of the photon ring and consequently the shape and size of the black hole shadow. Such effects could potentially be detected by very high-resolution observations, such as those of the Event Horizon Telescope (EHT) and the GRAVITY experiment.

Gravitational lensing effects also respond to the presence of charges. The deflection of light, the formation of multiple images, the distortion of the Einstein ring, and the angular position of gravitational arcs can be significantly altered when the spacetime metric contains both electric and magnetic charges. In this context, observational programs such as LSST, Euclid and the Nancy Grace Roman Space Telescope can in principle provide valuable clues to such models.

Another relevant aspect arises from the possibility—predicted by several grand unification theories—that magnetic monopoles exist. Although they have not yet been experimentally detected, if they do exist, it is reasonable to expect that black holes could capture them and thereby acquire magnetic charge. This would not only modify the spacetime metric but could also directly affect the dynamics of charged particles in the surrounding environment, potentially influencing the structure of relativistic jets and accretion disks. In this context, observations of jets (e.g., with ALMA, Chandra, and others), in combination with general relativistic magnetohydrodynamic (GRMHD) modeling, can be used to investigate possible observational signatures associated with such effects.

Finally, the direct search for magnetic monopoles remains ongoing in experiments such as MoEDAL (CERN), IceCube, ANTARES, and Super-Kamiokande. If detected, the interaction of these objects with black holes could open up new possibilities in both astrophysics and particle physics, including the formation of dyonic black holes or the emission of monopoles in extreme processes.

Although there is currently no experimental evidence for the existence of magnetically charged black holes—or even of monopoles—advances in ultra-high-precision observations, such as those anticipated from LISA, upgrades to the EHT, and future telescopes, may eventually allow the detection of effects associated with small electric or magnetic charges in compact objects.


\acknowledgments{
MER thanks Conselho Nacional de Desenvolvimento Cient\'ifico e Tecnol\'ogico - CNPq, Brazil, for partial financial support. This study was financed in part by the Coordena\c{c}\~{a}o de Aperfei\c{c}oamento de Pessoal de N\'{i}vel Superior - Brasil (CAPES) - Finance Code 001.
FSNL acknowledges support from the Funda\c{c}\~{a}o para a Ci\^{e}ncia e a Tecnologia (FCT) Scientific Employment Stimulus contract with reference CEECINST/00032/2018, and funding through the research grants UIDB/04434/2020, UIDP/04434/2020 and PTDC/FIS-AST/0054/2021.



\end{document}